# Next-Generation Quantum Theory of Atoms in Molecules for the Ground and Excited State of the Ring-Opening of Cyclohexadiene (CHD)

Tian Tian[1], Tianlv Xu[1], Steven R. Kirk[1*], Michael Filatov[1,2] and Samantha Jenkins[1*]

[1]*Key Laboratory of Chemical Biology and Traditional Chinese Medicine Research and Key Laboratory of Resource Fine-Processing and Advanced Materials of Hunan Province of MOE, College of Chemistry and Chemical Engineering, Hunan Normal University, Changsha, Hunan 410081, China*
[2]*Department of Chemistry, Ulsan National Institute of Science and Technology (UNIST), 50 UNIST-gil, Ulsan 44919, Korea*

email: steven.kirk@cantab.net
email: mike.filatov@gmail.com
email: samanthajsuman@gmail.com

The factors underlying the experimentally observed branching ratio (70:30) of the (1,3-cyclohexadiene) CHD→HT (1,3,5-hexatriene) photochemical ring-opening reaction are investigated. The ring-opening reaction path is optimized by a high-level multi-reference DFT method and the density along the path is analyzed by the QTAIM and stress tensor methods. The performed density analysis suggests that, in both $S_1$ and $S_0$ electronic states, there exists an attractive interaction between the ends of the fissile σ-bond of CHD that steers the ring-opening reaction predominantly in the direction of restoration of the ring. It is suggested that opening of the ring and formation of the reaction product (HT) can only be achieved when there is a sufficient persistent nuclear momentum in the direction of stretching of the fissile bond. As this orientation of the nuclear momentum vector can be expected relatively rare during the dynamics, this explains the observed low quantum yield of the ring-opening reaction.

**Introduction**

The photochemical electrocyclic ring-opening reaction of 1,3-cyclohexadiene (CHD) to 1,3,5,-hexatriene (HT) is a prototypical reaction relevant for many photochemical ring-opening processes important for synthetic organic chemistry and biochemistry[1–5]. The mechanism of this reaction was extensively studied by both experimental spectroscopic techniques[1,6] and theoretical modeling[2–5,7–10]. According to the widely accepted mechanism of this reaction[1,3–5], the photoexcitation (a $\pi \rightarrow \pi^*$ transition) at the ground electronic state ($S_0$) equilibrium geometry brings the molecule to the first excited electronic state ($S_1$) ($1^1B$ state in the $C_2$ symmetry; the symmetry labels represent diabatic states), whereupon the nuclear wave-packet slides down the steep slope on the $S_1$ PES and the state's character changes from $1^1B$ to $2^1A$ (a $\pi \rightarrow \sigma^*$ transition). After that, the nuclear wave-packet continues to move on the $S_1$ PES and reaches an $S_1/S_0$ conical intersection (CI), where it switches to the $S_0$ state. Having switched to the $S_0$ state, the nuclear wave-packet splits into two branches, one traveling back to CHD and another propagating forward to HT. The latest gas phase experimental measurements yield 70:30 CHD:HT branching ratio[6], which is generally consistent with 60:40 ratio for the reaction in solution[11].

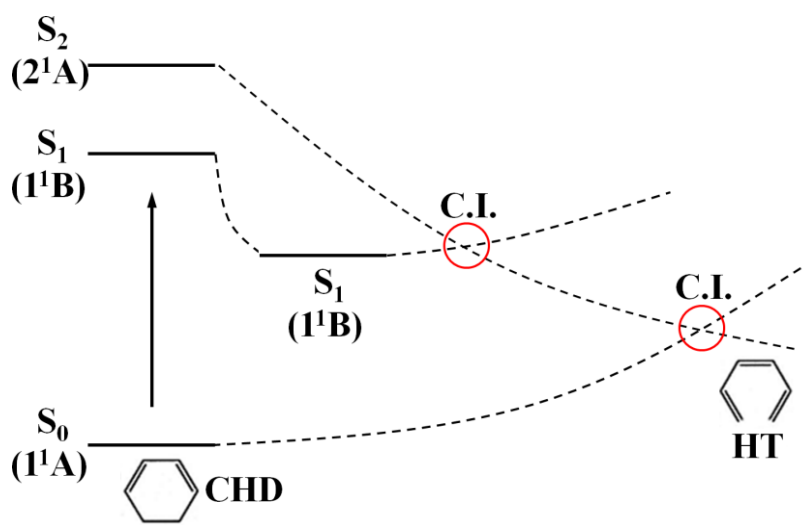

**Scheme 1.** Scheme of photochemical ring-opening reaction of CHD.

On the basis of CASSCF (and similar) calculations[2–5] it was concluded that upon the $S_1 \rightarrow S_0$ non-adiabatic population transfer through this conical intersection the wave-packet splits into two branches already on the $S_0$ surface. A 50:50 CHD:HT branching ratio was deduced on the basis of these calculations and quantum dynamics simulations on fitted PESs[2,7–9]. Early theoretical analysis of the CHD ring-opening reaction was heavily dominated by the CASSCF calculations, which predicted for the $S_1/S_0$ C.I. of CHD a bicyclic structure where the fissile C5-C6 bond and a bond with a neighboring atom form a nearly isosceles triangle[3–5,12].

The CASSCF calculations however, neglect the dynamic electron correlation, which has dramatic effect for the geometry of the $S_1/S_0$ conical intersection. The MSPT2 calculations resulted in a monocyclic geometry of the $S_1/S_0$ C.I.[12]. The subsequent TSH/MSPT2 NAMD simulations[10] yielded branching ratio (60:40) in a better agreement with the experiment.

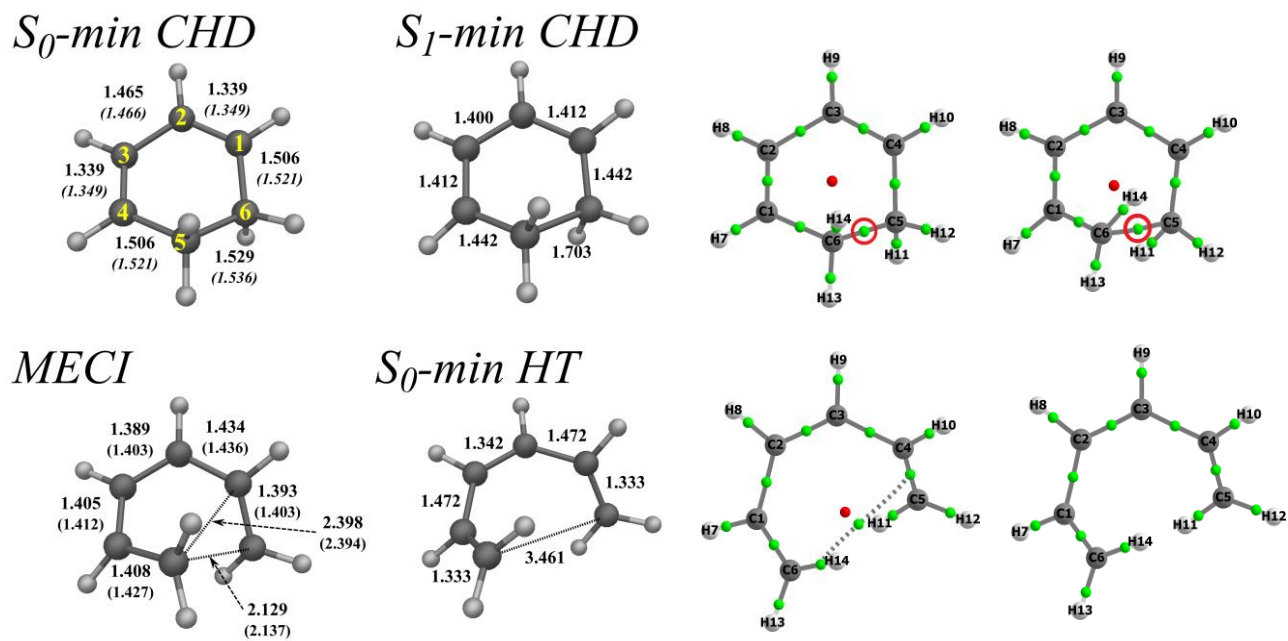

**Scheme 2.** Geometries of the ground $S_0$ and excited $S_1$ state minima and the $S_1/S_0$ MECI of CHD are shown in the left panel (gray scale). The values in parentheses show the experimental bond-lengths (italics) and bond-lengths obtained with the MSPT2 method (normal font) in Ref. 5. The corresponding molecular graphs (in color) for the CHD (FC-closed), CHD (C.I.) temporary state and HT (FC-open) configurations are shown in the right panel; with the ring-opening C5-C6 *BCP* being indicated by a red circle. Note, the C1-C5 bond in the grey-scale sketches corresponds to the C4---C6 *BCP* of the molecular graph. The undecorated red and green spheres indicate the positions of the ring critical points (*RCP*s) and bond critical points (*BCP*s) respectively.

Hence, a reasonable question to ask is what defines the CHD→HT photoreaction branching ratio? Are there electronic factors defining it, e.g., attractive bonding in the fissile C5-C6 bond, see **Scheme 2**, at the $S_1/S_0$ C.I., or is the branching ratio defined by purely dynamic factors, e.g., the momentum along the fissile C5-C6 bond?

To address these questions we have undertaken quantum chemical calculation of the minimum energy paths on the $S_1$ and $S_0$ PESs of the CHD ring-opening reaction. The computational method used in this work included the dynamic correlation from the outset; hence, predicting the correct $S_1/S_0$ C.I. with monocyclic geometry.

In this work, we shall investigate the ring-opening of the CHD along the MEPs in the $S_0$ and $S_1$ electronic states and analyze the resultant wave-functions with next-generation QTAIM that includes the bond-path framework set $\mathbb{B} = \{p,q,r\}$, see **Scheme 3**. Recently, the bond-path framework set $\mathbb{B}$ was used to investigate the $S_0$ and $S_1$ electronic states of fulvene[13]. The bond-path framework set $\mathbb{B} = \{p,q,r\}$ is constructed using the ellipticity ε as a scaling factor and comprises three paths associated with the least *p* and most *q* preferred directions of accumulation and the bond-path *r*. The ellipticity ε, was chosen as a scaling factor for $\mathbb{B}$ because it has already been found capable of differentiating between the $S_0$ and $S_1$ electronic states[13]. We will use next-generation QTAIM to attempt to investigate mechanism involved in the CHD:HT branching ratio.

## 2. Theory and Methods

*2.1 The QTAIM and stress tensor BCP properties; ellipticity ε, total local energy density $H(r_b)$ and the stress tensor eigenvalue $\lambda_{3\sigma}$*

We use QTAIM and the stress tensor analysis that utilizes higher derivatives of $\rho(\mathbf{r_b})$ in effect, acting as a 'magnifying lens' on the $\rho(\mathbf{r_b})$ derived properties of the wave-function. We will use QTAIM[14] to identify critical points in the total electronic charge density distribution $\rho(\mathbf{r})$ by analyzing the gradient vector field $\nabla\rho(\mathbf{r})$. These critical points can further be divided into four types of topologically stable critical points according to the set of ordered eigenvalues $\lambda_1 < \lambda_2 < \lambda_3$, with corresponding eigenvectors $\underline{e}_1$, $\underline{e}_2$, $\underline{e}_3$ of the Hessian matrix. In the limit that the forces on the nuclei become vanishingly small, an atomic interaction line[15] becomes a bond-path, although not necessarily a chemical bond[16]. The complete set of critical points together with the bond-paths of a molecule or cluster is referred to as the molecular graph.

In this investigation we have the closed-shell C5--C6 *BCP* and shared-shell C5-C6 *BCP* that comprise the CHD molecule, see **Scheme 2**. The bond notations "--" and "-" are used to distinguish instances where the Laplacian $\nabla^2\rho(\mathbf{r}) > 0$ for the closed-shell *BCP*s and $\nabla^2\rho(\mathbf{r}) < 0$ for shared-shell *BCP*s. Closed-shell BCPs may be divided into two categories: the strongest and weakest denoted by "--" and "---" corresponding to values of the total local energy density $H(\mathbf{r_b}) < 0$ and $H(\mathbf{r_b}) > 0$ respectively, see equation **(1)**. We include in this investigation an example of the weakest category of bond; the closed-shell C4---C6 *BCP* that corresponds to the C1-C5 bond in **Scheme 2**.

The least and most preferred directions of electron accumulation correspond to the $\underline{e}_1$ and $\underline{e}_2$ eigenvectors respectively[17–19] where the $\underline{e}_3$ eigenvector with associated eigenvalue $\lambda_3$, indicates the direction of the bond-path at the *BCP*. The ellipticity $\varepsilon = |\lambda_1|/|\lambda_2| - 1$, where $\lambda_1$ and $\lambda_2$ are the negative eigenvalues with corresponding eigenvectors $\underline{e}_1$ and $\underline{e}_2$ respectively that provide the relative accumulation of $\rho(\mathbf{r_b})$ in the two directions perpendicular to the bond-path. Recently, for the 11-cis retinal subjected to a torsion ±θ, we demonstrated that the $\underline{e}_2$ eigenvector of the torsional *BCP* corresponded to the preferred +θ direction of rotation as defined by the PES profile[20].

The stress tensor eigenvalue $\lambda_{3\sigma}$ is also associated with the bond-path and previously values of $\lambda_{3\sigma} < 0$ were found to be associated with transition-type behavior in biphenyl[21] and molecular motors[22] and indicated the bond critical point was close to rupturing and therefore we described as being *unstable*.

The total local energy density $H(\mathbf{r_b})$[23,24] is defined as:

$$H(\mathbf{r}_b) = G(\mathbf{r}_b) + V(\mathbf{r}_b), \tag{1}$$

where $G(\mathbf{r}_b)$ and $V(\mathbf{r}_b)$ are the local kinetic and potential energy densities at a *BCP*, defines a degree of covalent character: values of $H(\mathbf{r}_b) < 0$ for the closed-shell interaction, $\nabla^2\rho(\mathbf{r}_b) > 0$, indicates a *BCP* with a degree of covalent character and conversely a positive $H(\mathbf{r}_b) > 0$ reveals a lack of covalent character for the closed-shell *BCP*. A shared-shell *BCP* always possesses both $\nabla^2\rho(\mathbf{r}_b) < 0$ and $H(\mathbf{r}_b) < 0$ corresponding to short strong bonds e.g. C-C and C-H bonds.

*2.2 The QTAIM, bond-path framework set $\mathbb{B} = \{p,q,r\}$*

The bond-path length (BPL) is defined as the length of the path traced out by the $\underline{\mathbf{e}}_3$ eigenvector of the Hessian of the total charge density $\rho(\mathbf{r})$, passing through the *BCP*, along which $\rho(\mathbf{r})$ is locally maximal with respect to any neighboring paths. The bond-path curvature separating two bonded nuclei is defined as the dimensionless ratio:

$$(\text{BPL} - \text{GBL})/\text{GBL}, \tag{2}$$

Where the BPL is the associated bond-path length and the geometric bond length GBL is the inter-nuclear separation. The BPL often exceeds the GBL particularly for weak or strained bonds and unusual bonding environments[25]. For 3-D bond-paths, there are minor and major radii of curvature specified by the directions of $\underline{\mathbf{e}}_2$ and $\underline{\mathbf{e}}_1$ respectively.[26] In this investigation we suggest the involvement of the $\underline{\mathbf{e}}_3$ eigenvector also, in the form of a bond-path twist because earlier it was observed during calculations of the $\underline{\mathbf{e}}_1$ and $\underline{\mathbf{e}}_2$ eigenvectors at successive points along the bond-path that in some cases, these eigenvectors, both being perpendicular to the bond-path tracing eigenvector $\underline{\mathbf{e}}_3$, 'switched places'. We recently observed that the calculation of the vector tip path following the unscaled $\underline{\mathbf{e}}_1$ eigenvector would then show a large 'jump' as it swapped directions with the corresponding $\underline{\mathbf{e}}_2$ eigenvector[13]. This phenomenon indicated a location where the ellipticity $\varepsilon = 0$ due to degeneracies in the corresponding $\lambda_1$ and $\lambda_2$ eigenvalues, see the **Supplementary Materials S4** for a discussion on the choice of the ellipticity $\varepsilon$ as the scaling factor.

With *n* scaled eigenvector $\underline{\mathbf{e}}_2$ tip path points on the *q*-path where $\varepsilon_i$ = ellipticity at the $i^{\text{th}}$ bond-path point $\mathbf{r}_i$ on the bond-path *r*. It should be noted that the bond-path *r* is associated with the $\lambda_3$ eigenvalues of the $\underline{\mathbf{e}}_3$ eigenvector does not take into account differences in the $\lambda_1$ and $\lambda_2$ eigenvalues of the $\underline{\mathbf{e}}_1$ and $\underline{\mathbf{e}}_2$ eigenvectors. Analogously, for the $\underline{\mathbf{e}}_1$ tip path points we on the *p*-path where $\varepsilon_i$ = ellipticity at the $i^{\text{th}}$ bond-path point $\mathbf{r}_i$ on the bond-path *r*, where the $\mathbf{p}_i$ and $\mathbf{q}_i$ are defined by:

$$\mathbf{p}_i = \mathbf{r}_i + \varepsilon_i \underline{\mathbf{e}}_{1,i} \tag{3a}$$

$$\mathbf{q}_i = \mathbf{r}_i + \varepsilon_i \underline{\mathbf{e}}_{2,i} \tag{3b}$$

We referred to the next-generation QTAIM interpretation of the chemical bond as the *bond-path framework set*, denoted by $\mathbb{B}$, where $\mathbb{B} = \{p,q,r\}$ with the consequence that for the ground state a bond is comprised of three 'linkages'; *p*-, *q*- and *r*-paths associated with the $\underline{e_1}$, $\underline{e_2}$ and $\underline{e_3}$ eigenvectors, respectively.

The *p* and *q* parameters define eigenvector-following paths with lengths $\mathbb{H}^*$ and $\mathbb{H}$, see **Scheme 2**:

$$\mathbb{H}^* = \sum_{i=1}^{n-1} |\boldsymbol{p}_{i+1} - \boldsymbol{p}_i| \qquad (4a)$$

$$\mathbb{H} = \sum_{i=1}^{n-1} |\boldsymbol{q}_{i+1} - \boldsymbol{q}_i| \qquad (4b)$$

The lengths of the *eigenvector-following paths* $\mathbb{H}^*$ or $\mathbb{H}$ refers to the fact that the tips of the scaled $\underline{e_1}$ or $\underline{e_2}$ eigenvectors sweep out along the extent of the bond-path, defined by the $\underline{e_3}$ eigenvector, between two bonded nuclei connected by a bond-path. In the limit of vanishing ellipticity $\varepsilon = 0$, *for all* steps *i* along the bond-path then $\mathbb{H}$ = BPL.

From equation **(3a)** and equation **(3b)** we see for shared-shell *BCP*s, in the limit of the ellipticity $\varepsilon \approx 0$ i.e. corresponding to single bonds, we then have $p_i = q_i = r_i$ and therefore the value of the lengths $\mathbb{H}^*$ and $\mathbb{H}$ attain their lowest limit; the bond-path length (*r*) BPL. Conversely, higher values of the ellipticity ε, for instance, corresponding to double bonds will always result in values of $\mathbb{H}^*$ and $\mathbb{H}$ > BPL. Discussion on the uniqueness of the $\mathbb{H}^*$ and $\mathbb{H}$ is provided in the **Supplementary Materials S4**.

Analogous to the bond-path curvature, see equation **(2)**, we may define dimensionless, *fractional* versions of the eigenvector-following path with length $\mathbb{H}$ where several forms are possible and not limited to the following:

$$\mathbb{H}_f = (\mathbb{H} - \text{BPL})/\text{BPL} \qquad (5)$$

A similar expressions to equation **(5)** for $\mathbb{H}^*_f$ can be derived using the $\underline{e_1}$ eigenvector.

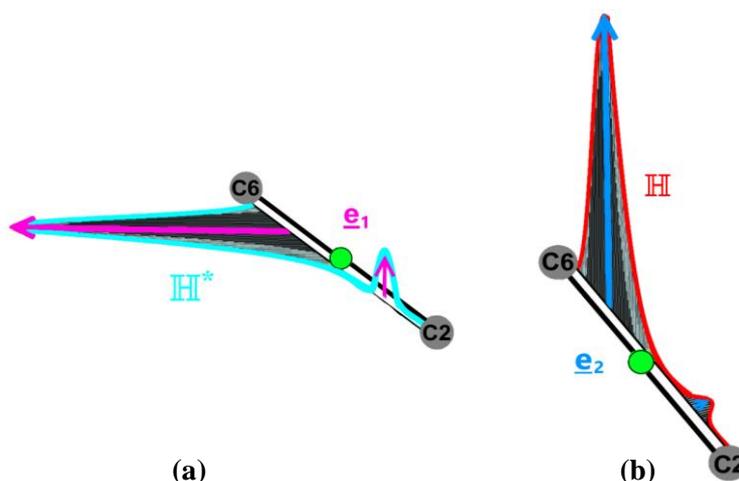

(a)      (b)

**Scheme 3.** The pale-blue line in sub-figure **(a)** represents the path, referred to as the eigenvector-following path with length $\mathbb{H}^*$, swept out by the tips of the scaled $\underline{e_1}$ eigenvectors, shown in magenta, defined by equation **(4a)**. The red

path in sub-figure (**b**) corresponds to $\mathbb{H}$, constructed from the path swept out by the tips of the scaled $\underline{e}_2$ eigenvectors, shown in mid-blue and is defined by equation (**4b**). The pale-blue and mid-blue arrows representing the $\underline{e}_1$ and $\underline{e}_2$ eigenvectors are scaled by the ellipticity ε respectively, where the vertical scales are exaggerated for visualization purposes. The green sphere indicates the position of a given *BCP*. Details of how to implement the calculation of the $\mathbb{H}^*$ and $\mathbb{H}$ are provided in the **Supplementary Materials S4**.

Previously, we considered the $S_0$ and $S_1$ electronic states of fulvene for the bond-path framework set $\mathbb{B} = \{(p_0, p_1), (q_0, q_1), r\}$ [13].

## 3. Computational Details

The potential energy surfaces of the $S_1$ and $S_0$ states of CHD along the ring-opening reaction coordinate were investigated using the state-interaction state-averaged REKS (SI-SA-REKS or SSR, for brevity) method[27]. The SSR method employs ensemble density functional theory to describe the non-dynamic electron correlation occurring due to multi-reference character of the $S_0$ state and to obtain excitation energies from a variational time-independent formalism. By contrast to the popular state-averaged CASSCF (SA-CASSCF) methodology, the SSR method includes the dynamic electron correlation from the outset; thus providing very accurate description of the PESs of the $S_1$ and $S_0$ states and conical intersections between the states[28,29].

In this work, the SSR method is employed in connection with the ωPBEh range-separated hybrid density functional[30] and the 6-31G* basis set[31]. The geometries of the $S_1$ and $S_0$ state minima, $S_1/S_0$ conical intersections and the minimum energy paths (MEPs) were optimized using the SSR analytical energy derivatives formalism described in Ref.[32]. The SSR computations were carried out using the beta-testing version of the TeraChem® program (v1.92P, release 7f19a3bb8334)[33–38].

Building the $S_1$ and $S_0$ MEPs was carried out as follows: First, the geometries of the $S_1$ and $S_0$ minima were optimized for CHD and cZc-HT using the DL-FIND module[39] interfaced with TeraChem. Then, the $S_1/S_0$ conical intersection geometry was optimized using the CIOpt program[40] receiving the SSR energies and gradients from TeraChem output. Having obtained the $S_1$ and $S_0$ equilibrium geometries and the MECI geometry, the MEPs connecting these critical points were optimized using the nudged elastic band (NEB) method[41] with fixed endpoints as implemented in DL-FIND. The MEPs comprise the following legs: (1) CHD-$S_0$min – $S_1/S_0$ MECI, (2) CHD-$S_1$FC – $S_1/S_0$ MECI, (3) $S_1/S_0$ MECI – cZc-HT-$S_0$min and (4) $S_1/S_0$ MECI – cZc-HT-$S_1$FC; the MEPs (1) and (3) are on the $S_0$ PES, the MEPs (2) and (4) are on the $S_1$ PES. The MEPs (1) and (2) comprise 20 NEB images, the MEPs (3) and (4) comprise 40 images each. At each point of the MEPs the relaxed density matrix for the respective state was calculated and analyzed with QTAIM and the stress tensor using the AIMAll[42] suite on each wave-function obtained in the previous step.

# 4. Results and discussions

*4.1. Analysis of the $S_1$ and $S_0$ states minimum energy pathways of the CHD ring-opening reaction*

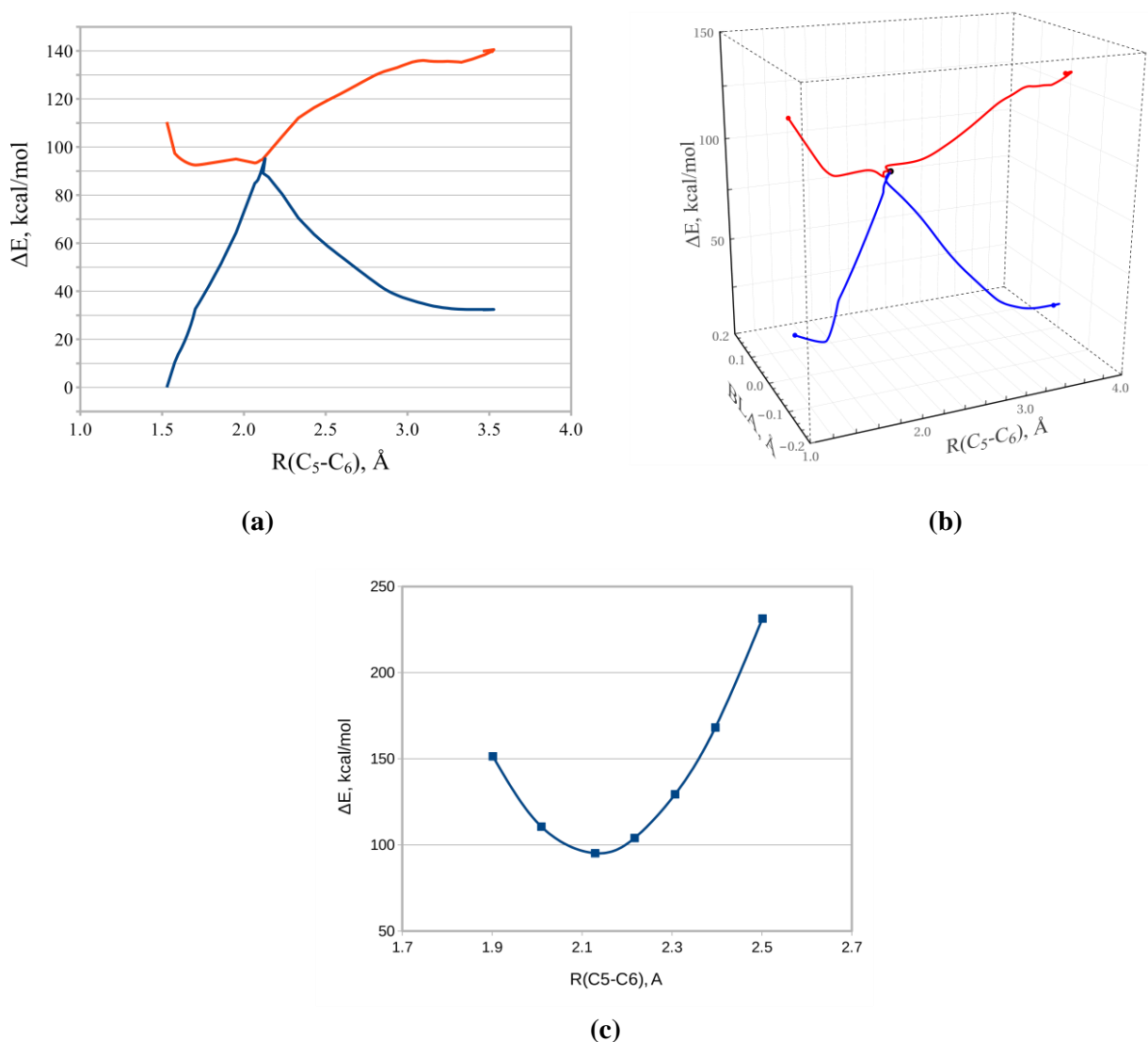

**Figure 1**. The variation of the 2-D and 3-D relative energy $\Delta E$ for the $S_0$(blue) and $S_1$(red) electronic states of the CHD→HT photoreaction with R(C5-C6) in Ångstrom are presented in the sub-figure **(a-b)** respectively. Additional relative energies $\Delta E$ of additional C.I. points along the $S_1/S_0$ C.I. seam of CHD are presented in sub-figure **(c)**.

The geometries corresponding to the local minima on the $S_1$ and $S_0$ PESs of CHD and the $S_1/S_0$ MECI optimized in this work using the SSR-ωPBEh/6-31G* method are shown in **Scheme 1** and **Scheme 2**. The $S_0$-min CHD geometry obtained is in a reasonable agreement with the experimental geometry available in the literature[43]. The $S_1/S_0$ MECI geometry is in good agreement with the recent MSPT2 geometry from ref.[12].

Profiles of the $S_0$ and $S_1$ PESs along the MEPs of the ring-opening of CHD demonstrate that the photoexcitation at the $S_0$ equilibrium geometry of CHD is achieved by the π→π* one-electron transition, which weakens the double bonds of the CHD ring, see **Figure 1**, **Scheme 1** and **Scheme 2**. Simultaneously with the π-bonds, the C5-C6 σ-bond is slightly weakened by the excitation. At the $C_2$ geometry, the σ-bonding orbital is mixed with the π-type HOMO and depletion of the electron density in the HOMO leads to weakening of the σ-bond. The excitation leads to stretching of the former π-bonds and the C5-C6 σ-bond of CHD and the $S_1$ MEP slides down to the $S_1$ minimum, which occurs at a $C_2$ geometry with a stretched C5-C6 bond and π-bonds of CHD. The profile of the MEP where the $S_1$ and $S_0$ relative energies ΔE are shown as functions of the C5-C6 bond length and the bond-length-alternation (BLA) distortion is presented the 3-D plot in **Figure 1(b)**. At the beginning of the $S_1$ MEP, the BLA coordinate undergoes a rapid change from a positive value (double bonds shorter than single bonds) to a near zero value (double and single bonds are nearly equal in length).

After the $S_1$ minimum, the MEP follows in the direction of the MECI point, which occurs at a geometry distorted away from the $C_2$-symmetry. The MECI geometry corresponds to a six-membered ring, where the C5-C6 bond is stretched to 2.129 Å. This agrees well with the geometry obtained by the MSPT2 method. The earlier CASSCF calculations yielded a five-membered ring geometry corresponding to the MECI, with the C1-C5 bond (referred to as the C4---C6 *BCP* in this work, see **Scheme 2**) shorter than the C5-C6 bond. The CASSCF method, however, neglects the dynamic electron correlation. The inclusion of the dynamic correlation in the SSR (and in the MSPT2) method has considerable implications for the geometry of the $S_1/S_0$ MECI of CHD.

Having crossed the MECI point, the $S_0$ MEP continues to the cZc-HT $S_0$ minimum without any hindrance. Similarly, the $S_0$ MEP leading to the original CHD conformation is barrierless and steep. There is however, an indication that the MECI→HT stretch of the $S_0$ MEP first goes in the direction of shorter C5-C6 bond and then turns to stretching of the bond, see **Figure 1(a)**. This implies the existence of an attractive interaction between the C5 and C6 atoms in the vicinity of the MECI geometry.

To investigate whether the attractive interaction is present at C.I. geometries with a stretched C5-C6 bond-length, a number of *additional* C.I.s along the $S_1/S_0$ seam were optimized with the bond-length varying in the range of 1.9 Å to 2.5 Å. The relative energies ΔE of the C.I.s (with respect to CHD $S_0$-min) are presented in **Figure 1(c)**. The C.I. points studied here cover sufficiently large energy range; up to 150 kcal/mol above the MECI point. The presence of an attraction between C5 and C6 atoms for the higher energy C.I.s implies that the C5-C6 bond is central to controlling the dynamics of CHD ring-opening. Therefore, the presence of the C5-C6 bonding interaction for stretched bond-lengths should steer the reaction toward CHD and opening of the ring can occur if there is a sufficiently large and persistent momentum pointing in the direction of the stretching of the C5-C6 bond.

*4.2. A QTAIM bond-path analysis of the CHD ring-opening reaction*

In this section we will refer to the MEPs outlined in section 4.1 in the context of the complete 3-D bond-path framework set $\mathbb{B} = \{(\boldsymbol{p_0},\boldsymbol{p_1}),(\boldsymbol{q_0},\boldsymbol{q_1}),\boldsymbol{r}\}$ and the bond-path ellipticity ε profiles to quantify the increase in closed-shell *BCP* character of the shared-shell C5-C6 *BCP* as the ring-open reaction proceeds, see **Figure 2** and **Figure 3** respectively.

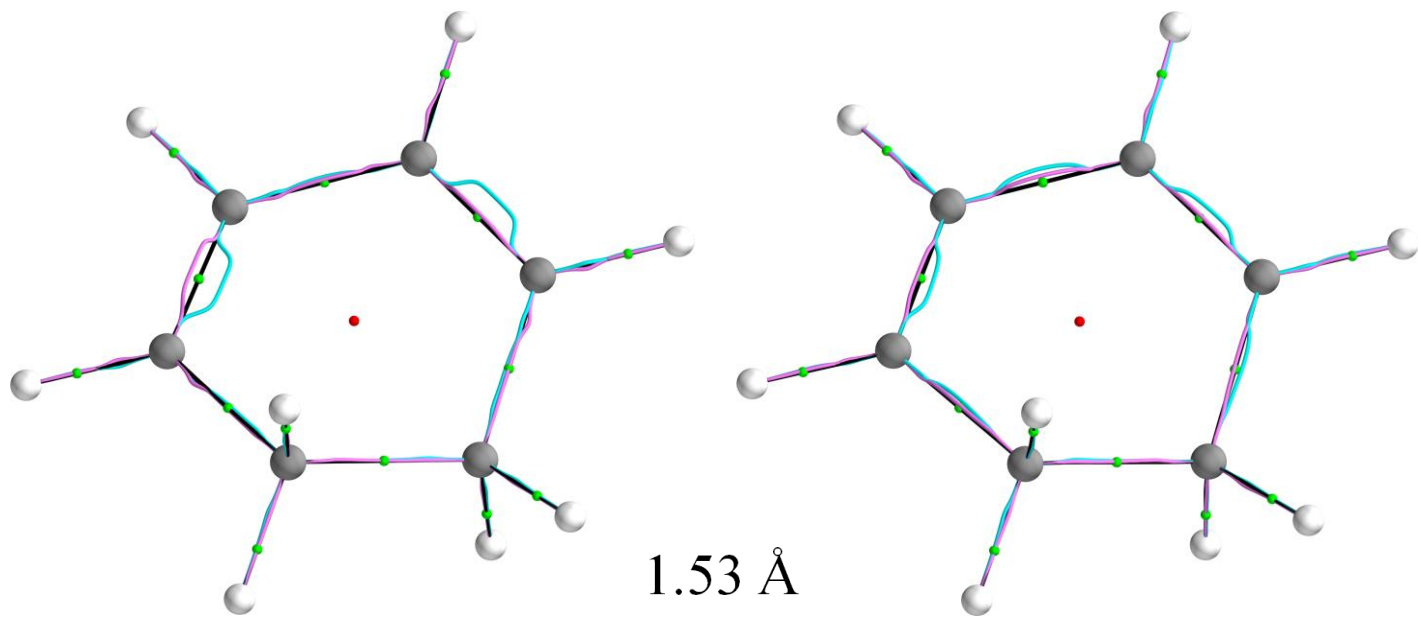

(a) 1.53 Å

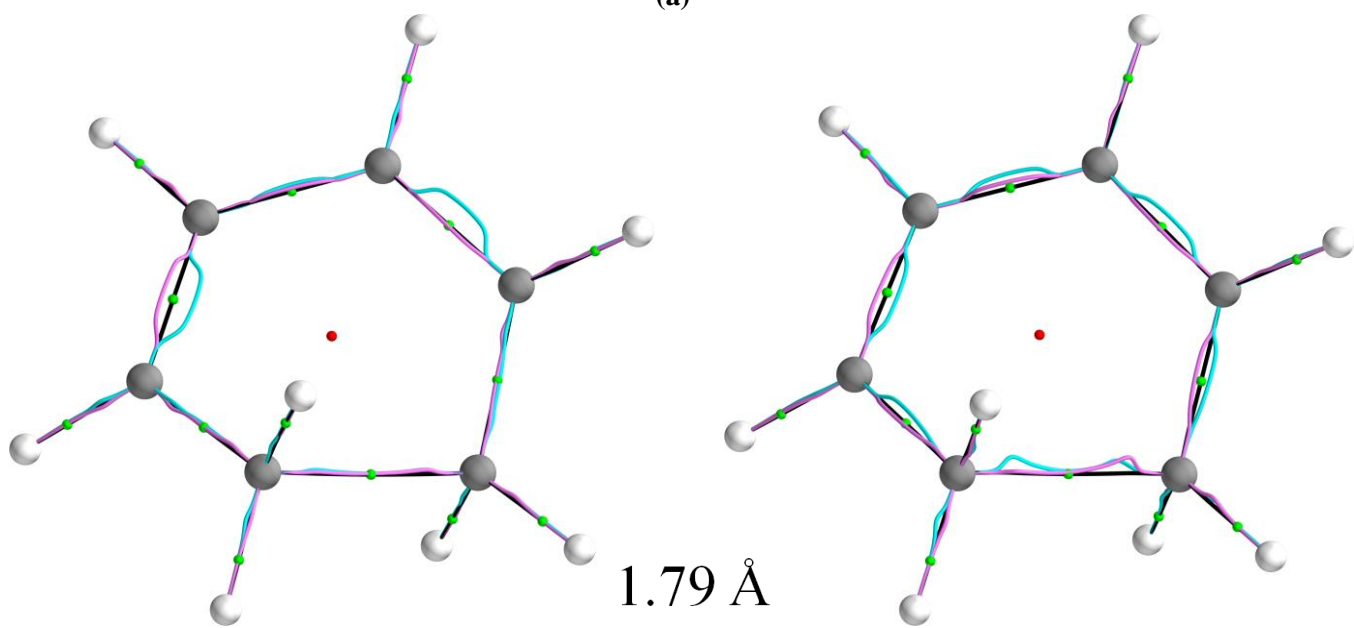

(b) 1.79 Å

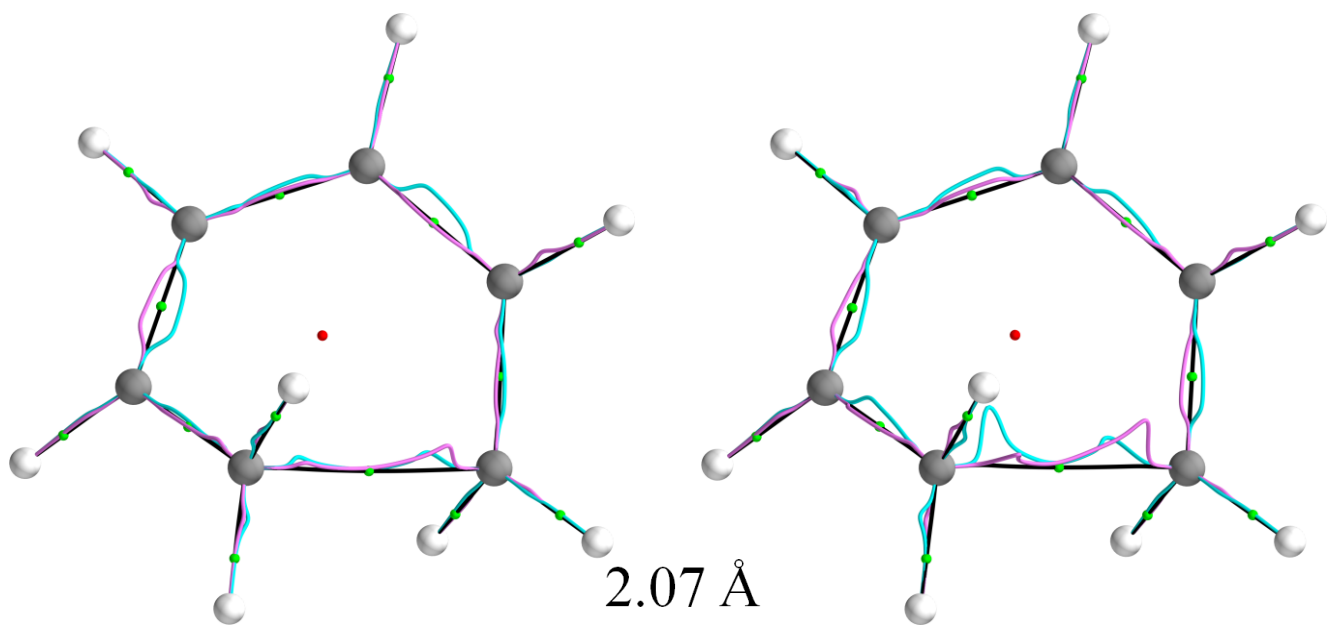

(c) 2.07 Å

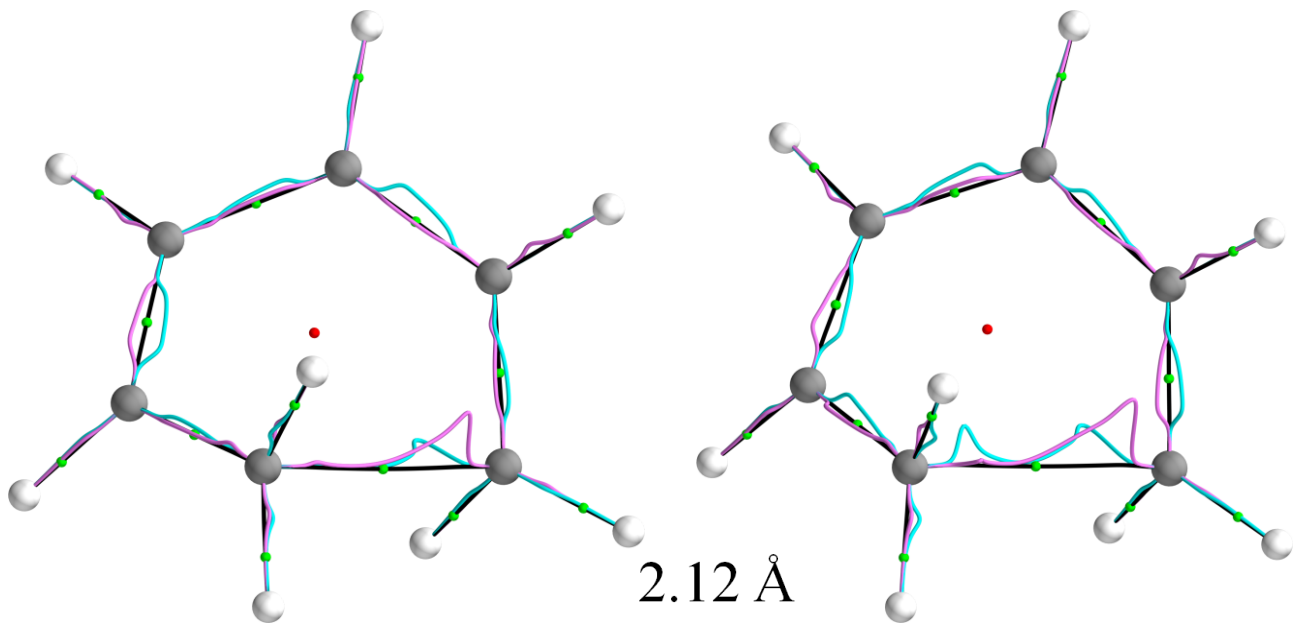

2.12 Å

(d)

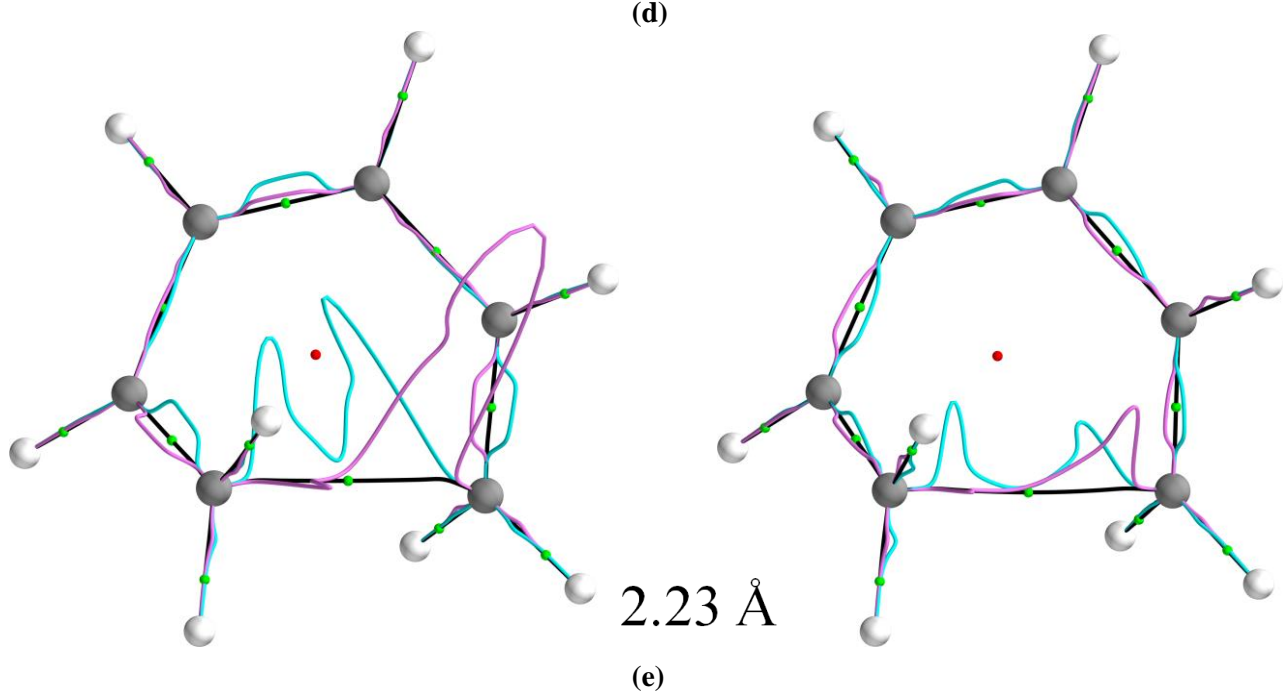

2.23 Å

(e)

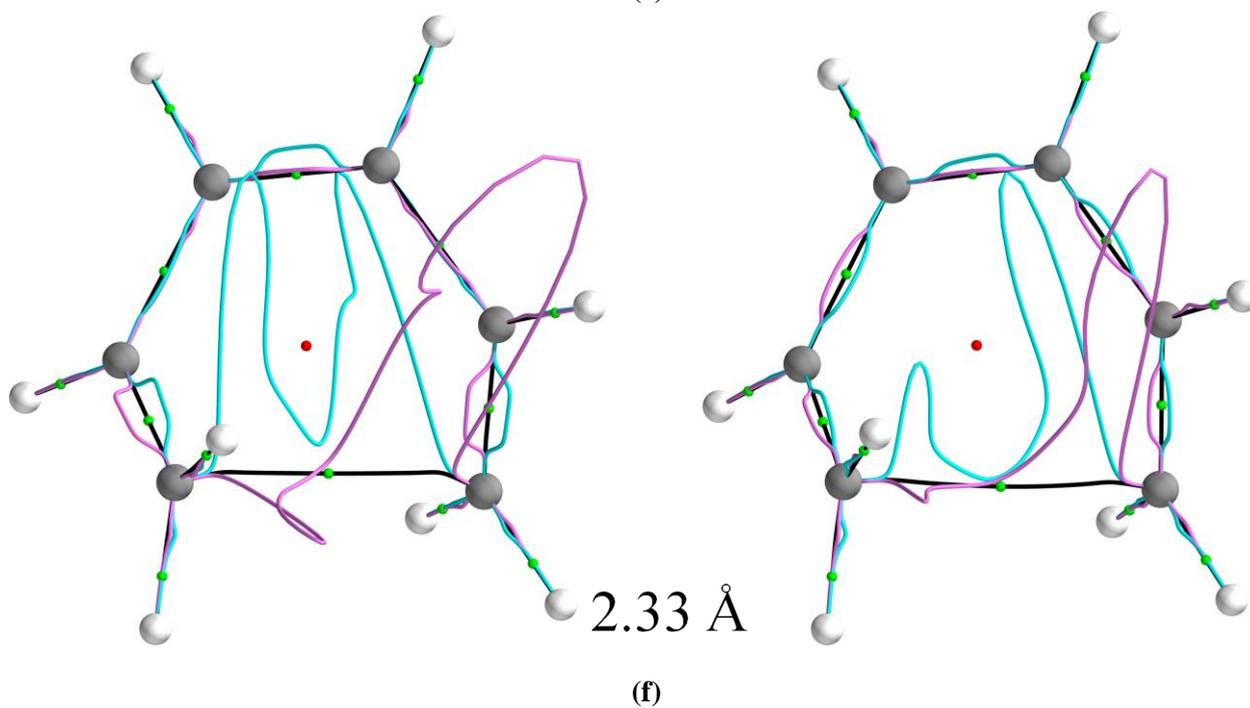

2.33 Å

(f)

**Figure 2.** The $p_0$- and $q_0$-paths for the $S_0$ state (left panel) and corresponding $p_1$- and $q_1$-paths $S_1$ state (right panel) electronic states of the C5-C6 *BCP* bond-path for the CHD→HT photoreaction molecular graph subjected to the bond-length R(C5-C6) distortion for 1.53 Å, 1.79 Å, 2.07 Å, 2.12 Å, 2.23 Å and 2.33 Å are presented in sub-figures **(a)-(f)** respectively. The undecorated green and red spheres indicate the positions of the bond critical points (*BCP*s) and ring critical points (*RCP*), see also **Figure 1** and **Scheme 2** for the atom numbering scheme.

Examination of the $p_0$- and $q_0$-paths and $p_1$- and $q_1$-paths for the $S_0$ and $S_1$ states respectively indicates that the reaction at CHD commences with a mix of double and single bond character bonds. Double and single bond character is indicated by the presence of the ($p_0$,$q_0$) or ($p_1$,$q_1$)-paths that display larger or smaller deviations around the *BCP* associated with the bond-path; note, the ($p_0$,$q_0$) or ($p_1$,$q_1$)-paths may be obscured by being normal to the plane of the molecule, see **Figure 2**.

The weakening of the double bonds of the CHD ring due to the CHD→HT photoreaction is evident from the smaller deviations of the ($p_1$,$q_1$)-paths compared with the ($p_0$,$q_0$)-paths, hence lower ellipticity ε values, away from the C3-C4 *BCP* and C1-C2 *BCP* bond-paths (*r*), shown in the left and right panels of **Figure 2(a)**, also see **Scheme 2**. The photoexcitation leads to stretching of the bond-path (*r*) of the C5-C6 *BCP* and results in longer ($p_1$,$q_1$)-paths than ($p_0$,$q_0$)-paths as the $S_1$ MEP slides down to the $S_1$ minimum; simultaneously it leads to the creation of higher ellipticity ε (π-bonds) of CHD ring.

We notice that after the MECI point (R(C5-C6)= 2.12Å) the ($p_0$,$q_0$) and ($p_1$,$q_1$)-paths associated with the C5-C6/C5--C6 *BCP* drastically stretches signifying the transition in chemical character from the (C5-C6 σ-bond) shared-shell C5-C6 *BCP* to the closed-shell C5--C6 *BCP*. In addition, after the MECI point the $q_0$- and $q_1$-paths (indicate the most preferred direction of motion of $\rho(\mathbf{r_b})$) are clearly directed normally to the plane of the CHD ring and not towards the *RCP* (undecorated red sphere at the centre of the CHD ring) that would be the case for an unstable closed-shell *BCP*. The eventual direction of rupture of the closed-shell C5--C6 *BCP* is indicated by the $q_0$- and $q_1$-paths.

Before the MECI point is reached, the ellipticity ε profiles indicate that the values of the ellipticity ε are higher for the $S_1$ than for the $S_0$ state, see **Figure 3(a-c)**; the converse being true after the MECI point, see **Figure 3(d-f)**. The ellipticity ε profiles demonstrate that throughout the duration of the CHD ring-opening there is an increasing tendency for the $\rho(\mathbf{r_b})$ to accumulate towards the C5 *NCP* and C6 *NCP*. This demonstrates an increase in closed-shell *BCP* character and therefore dominance of the positive $\lambda_3$ eigenvalue resulting in a value of the Laplacian $\nabla^2\rho(\mathbf{r_b}) > 0$, where $\nabla^2\rho(\mathbf{r_b}) = \lambda_1 + \lambda_2 + \lambda_3$ and is entirely consistent with the behavior of closed-shell *BCP*s. Note that the Hessian eigenvalues are ordered as $\lambda_1 < \lambda_2 < \lambda_3$, and we always have $\lambda_1 < 0$ and $\lambda_2 < 0$. Therefore, the increase in the magnitude of the peaks in the ($p_0$,$q_0$) and ($p_1$,$q_1$)-paths towards C5 *NCP* and C6 *NCP* displays the variation in closed-shell *BCP* character along the bond-path of the C5-C6/C5--C6 *BCP*.

In the next section, we explore the mixture of shared-shell *BCP* character of the closed-shell C5--C6 *BCP*.

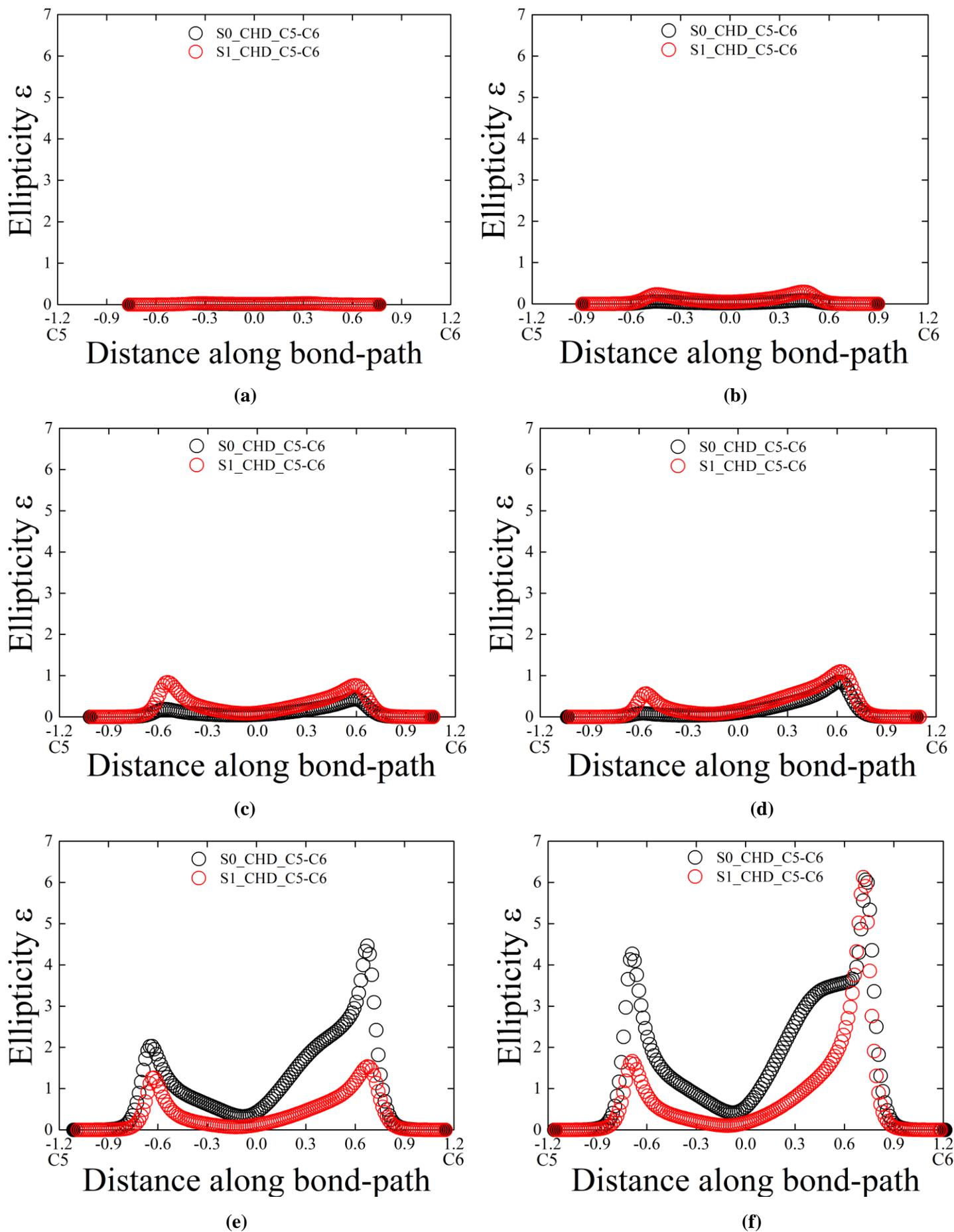

**Figure 3.** The ellipticity ε profiles along the bond-path of the C5-C6 *BCP* for the CHD→HT photoreaction molecular graph corresponding to bond length R(C5-C6) distortions of 1.53 Å, 1.79 Å, 2.07 Å, 2.12 Å, 2.23 Å and 2.33 Å in the $S_0$ and $S_1$ electronic states are presented in sub-figures **(a)**-**(f)**, see **Scheme 2** for the atom labelling scheme.

## 4.3. A QTAIM and stress tensor BCP analysis of the CHD ring-opening reaction

In section 4.2 we tracked the increase in closed-shell *BCP* character of the shared-shell C5-C6 *BCP* from the start of the CHD→HT photoreaction. In this section we mainly focus on quantifying the mixture of shared-shell C5-C6 *BCP* (C5-C6 σ-bond) character of the closed-shell C5--C6 *BCP* as measured by the presence of values of the total local energy density $H(\mathbf{r_b}) < 0$. We also include the temporary C4---C6 *BCP*, see **Figure 4**, which is referred to as the C1-C5 bond in the grey scale sketches of the left panel of **Scheme 2**. We also consider the changes to the topological stability from values of the stress tensor eigenvalue $\lambda_{3\sigma} < 0$ and increase in the ellipticity ε that is known to indicate a decrease in stability for closed-shell *BCP*s, see **Figure 4(b)** and **Figure 4(c)**, respectively. To display the variations of the $H(\mathbf{r_b})$, $\lambda_{3\sigma}$ and the ellipticity ε values with the bond-path length (BPL) over both the shared-shell *BCP* and closed-shell *BCP* regimes with clarity and continuity we split the results across overlapping ranges of bond-path length values, (BPL = 1.5 Å to 2.2 Å) and (BPL = 2.0 Å to 4.2 Å) in the left and right panels of **Figure 4(a-c)**.

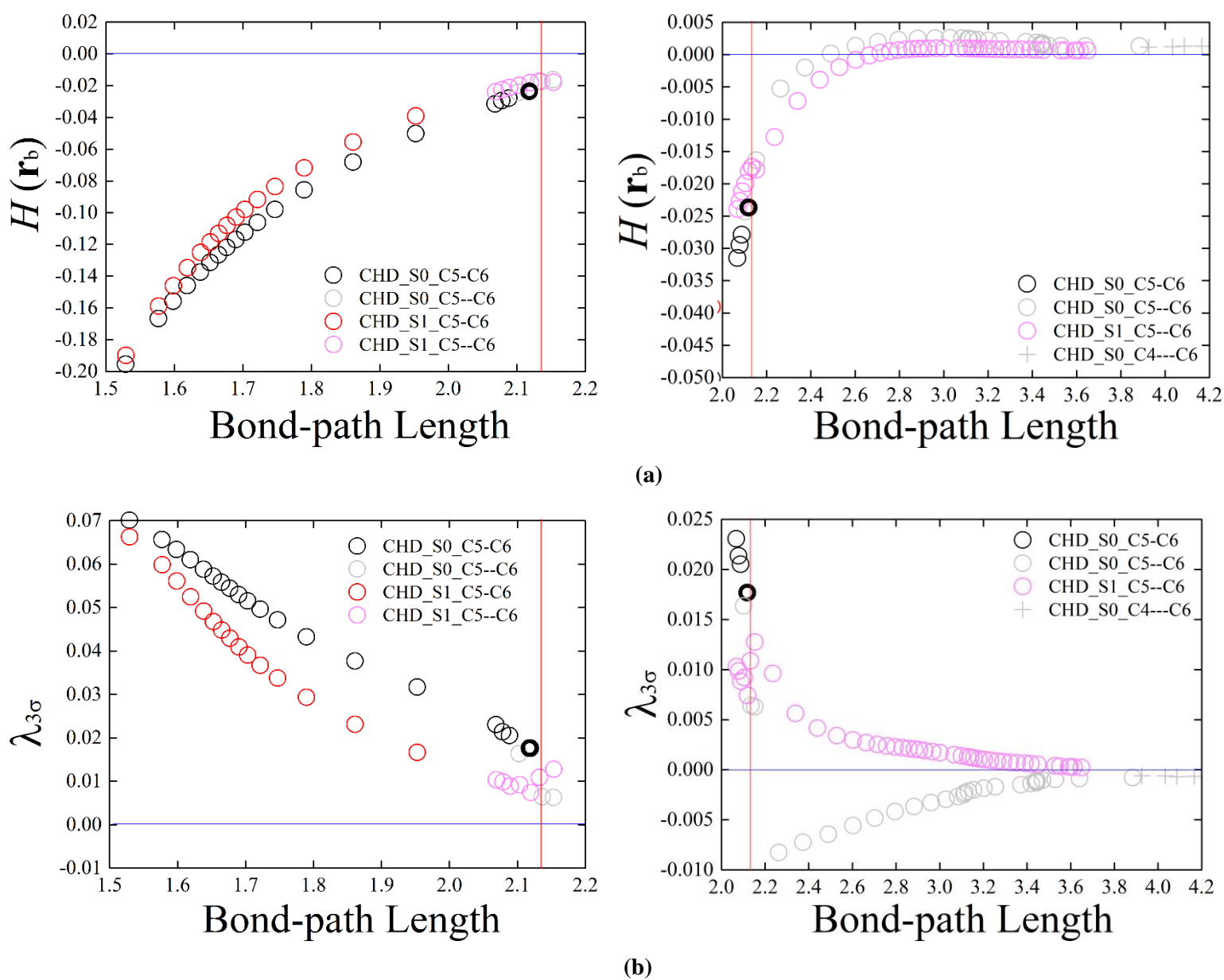

(a)

(b)

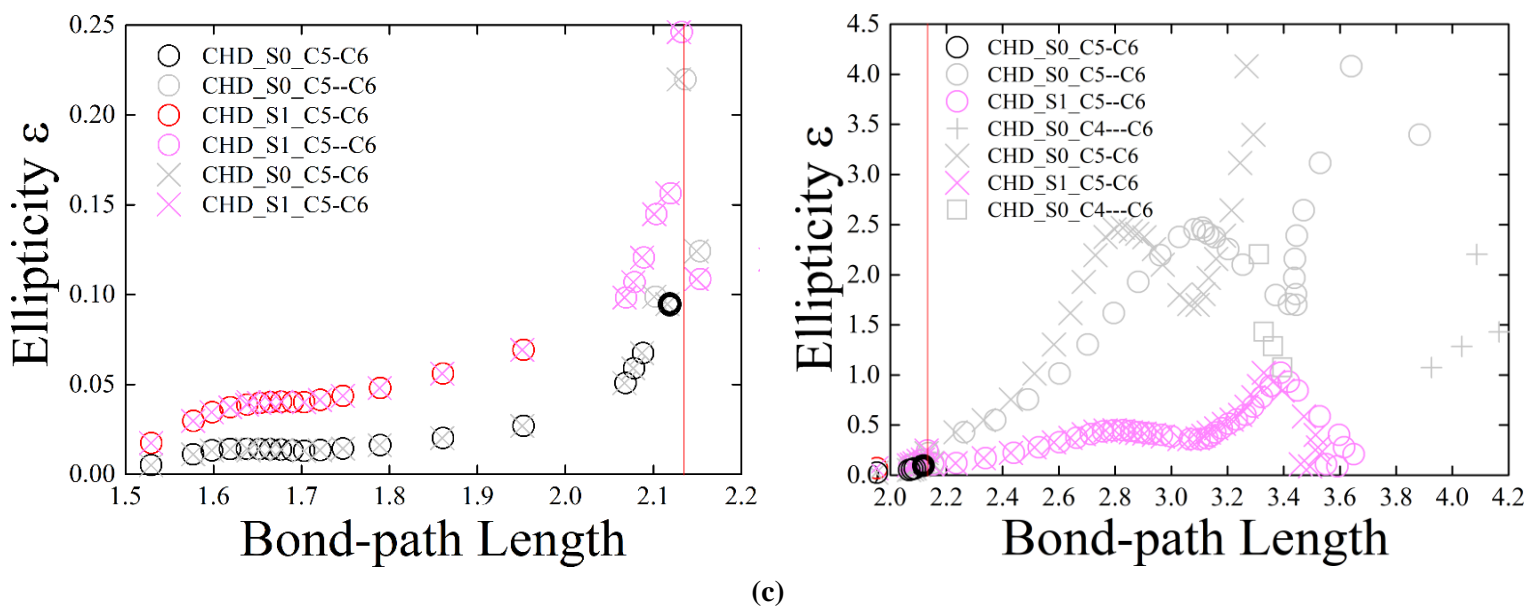

(c)

**Figure 4**. The variation of the local total energy density $H(\mathbf{r_b})$, stress tensor eigenvalue $\lambda_{3\sigma}$ and ellipticity ε for $S_0$ (grey) and $S_1$ (pale-magenta) with the bond-path length (BPL) in Å of the closed-shell C5--C6 *BCP* (circle) and C4---C6 *BCP* (cross) for the of the CHD→HT photoreaction molecular graph are presented in sub-figures **(a-c)** respectively. The horizontal blue line and vertical red lines correspond to values of $H(\mathbf{r_b}) < 0$ and the position of the conical intersection (C.I.), respectively see **Scheme 2** and **Figure 1(a-b)**. Note, for the $S_0$ state the bold black circle corresponds to a point on the MEP beyond the C.I. Values of $H(\mathbf{r_b}) < 0$ and the MECI point are highlighted by the horizontal blue line and the vertical red line respectively. The shared-shell *BCP*s (denoted by "-") and closed-shell *BCP*s where $H(\mathbf{r_b}) < 0$ and $H(\mathbf{r_b}) > 0$ are denoted by "--" and "---" respectively in the legends.

At the start of the reaction the fissile C5-C6 bond is a shared-shell C5-C6 *BCP*, indicated by the black and red circles for the $S_0$ and $S_1$ states respectively, see **Scheme 1**, **Scheme 2** and the left panels of **Figure 4(a-c)**. Before the MECI the C5-C6 *BCP* is stronger in the $S_0$ than the $S_1$ state as indicated by more negative values $H(\mathbf{r_b})$ and more topologically stable as determined by more positive values of $\lambda_{3\sigma}$ associated with the $S_0$ state. Also before the MECI, the ellipticity ε is lower in the $S_0$ than the $S_1$ state consistent with stronger shared-shell *BCP*s. In the $S_0$ state there is an oscillation in the chemical nature of the C5-C6 bond between the shared-shell C5-C6 *BCP* and closed-shell C5--C6 *BCP*, see the left panel of **Figure 4(a)**. The bold black circle (a shared-shell C5-C6 *BCP*) corresponds to a point on the $S_0$ MEP beyond the MECI but appears to occur before the MECI due to a contraction of the C5-C6 *BCP* bond-path resulting in transformation of the closed-shell C5--C6 *BCP* → shared-shell C5-C6 *BCP*. This oscillation provides evidence that the MECI→HT stretch of the $S_0$ MEP first goes in the direction of shortening the closed-shell C5--C6 *BCP* back to stronger shared-shell *BCP* before continuing with the stretching of the bond-path of the closed-shell C5--C6 *BCP*. We also notice that for $S_0$ the shared-shell C5-C6 *BCP*, indicated by the bold black circle, has a more positive value of $\lambda_{3\sigma}$ than the neighboring closed-shell *BCP* (grey circle) that possesses a shorter BPL, see **Figure 4(a)**. This indicates that the highlighted (bold black circle) shared-shell C5-C6 *BCP* is more topologically stable than the neighboring closed-shell C5--C6 *BCP*.

For a value of R(C5-C6) = 2.12 Å the $S_0$ state corresponds to the shared-shell C5--C6 *BCP* in contrast to the closed-shell C5--C6 *BCP* for the $S_1$ state, see the left and right panels of **Figure 2(d)** respectively, see also the **Supplementary Materials S5**. This demonstrates that for the $S_0$ state there is an attractive interaction between the C5 *NCP* and C6 *NCP* in the vicinity of the MECI point and an attraction beyond the MECI geometry up to at least 2.5 Å, see **Figure 5**.

Differences in the variation with the ellipticity ε of the C5 *NCP* and C6 *NCP* separation ($S_0$ and $S_1$ indicated by pale-magenta and grey 'X' respectively), compared with the bond-path length demonstrate bond stretching before and after the MECI point, see the left and right panels of **Figure 4(c)** respectively. The stretching of the C5-C6 *BCP* bond-path is particularly large for the $S_0$ state and much larger than for the $S_1$ state, after the MECI point, explaining why the MEP continues to the cZc-HT $S_0$ minimum without hindrance, see the right panel of **Figure 4(c)**.

In the $S_0$ state the shared-shell C5-C6 *BCP* persists further along the MEP than in the $S_1$ state, see the right panels of **Figure 4(a-c)**. Along the MEP in the $S_1$ state values of the Laplacian $\nabla^2\rho(\mathbf{r}_b) < 0$ corresponding to the shared-shell C5-C6 *BCP* transforms smoothly to values of $\nabla^2\rho(\mathbf{r}_b) > 0$, which correspond to the closed-shell C5--C6 *BCP*, indicated by the pale-magenta circles in **Figure 4(a-c)**.

The mixed chemical character of the C5-C6 σ-bond discussed in section 4.1 was explained in terms of the σ-bonding orbital is mixing with the π-type HOMO and depletion of the electron density in the π-orbital resulting in the weakening and stretching of the C5-C6 σ-bond. The stretching results in the transition from the shared-shell C5-C6 *BCP* to the closed-shell C5--C6 *BCP* before the MECI point, see **Figure 4**.

We can quantify the mixed chemical character in terms of the existence of values of $H(\mathbf{r}_b) < 0$ for the closed-shell C5--C6 *BCP*, that indicates a degree of coupling between the C5--C6 *BCP* and neighboring shared-shell *BCP*s, see **Figure 4(a)**. As the ring-opening reaction proceeds, the closed-shell *BCP*s retain a degree of shared-shell *BCP* character, i.e. $H(\mathbf{r}_b) < 0$, beyond the MECI point with lower values for the $S_0$ than for the $S_1$ state, see the right panel of **Figure 4(a)**. This is consistent with the finding that after the MECI point, the MEP continues to the cZc-HT $S_0$ minimum without hindrance. In addition, after the MECI point the values for $\lambda_{3\sigma}$ in the $S_0$ state become less *negative* despite the stretching of the C5--C6 *BCP* bond-path, conversely the values for $\lambda_{3\sigma}$ in the $S_1$ state become less positive, see **Figure 4(b)**. For the $S_0$ state the C4---C6 *BCP* (referred to as the C1-C5 bond in left panel of **Scheme 2**) temporarily exists and for all values the C4---C6 *BCP* is weak (since $H(\mathbf{r}_b) > 0$) and topologically unstable, $\lambda_{3\sigma} < 0$, see the right panels of **Figure 4(a)** and **Figure 4(b)** respectively.

A number of *additional* C.I.s along the $S_1/S_0$ seam were optimized with the bond-path C5-C6 *BCP* in the range of 1.9 Å to 2.5 Å, see **Figure 5**. The C.I. points investigated cover sufficiently large energy range; up to 150 kcal/mol above the MECI point, see **Figure 1(c)**. The results presented for the additional points along the C.I. seam show that for bond-path lengths of 2.5 Å the closed-shell C5--C6 *BCP* possesses a degree of covalent character i.e. $H(\mathbf{r}_b) < 0$ and therefore is "sticky" and resistant to moving towards the HT product until $H(\mathbf{r}_b) > 0$.

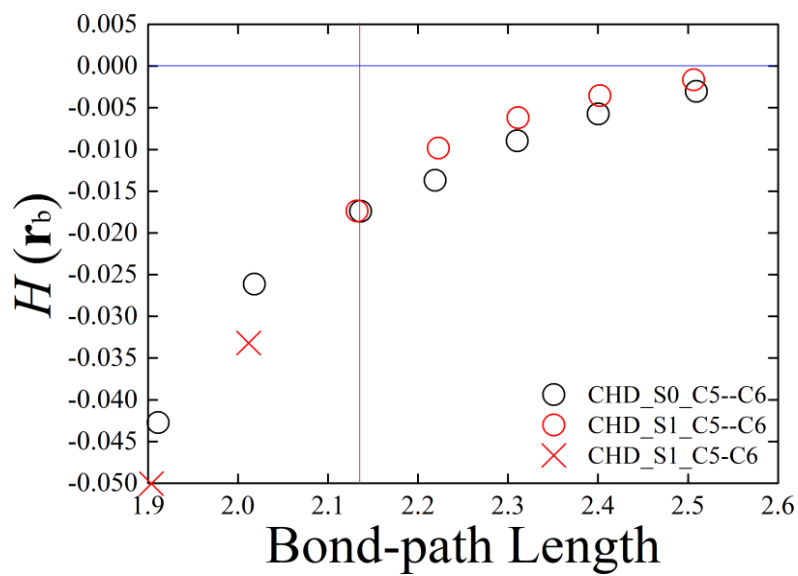

**Figure 5.** The variation of local total energy density $H(\mathbf{r_b})$ for $S_0$ (black) and $S_1$ (red) with the bond-path length (Å) of the closed-shell C5--C6 *BCP* (circle) and shared-shell C5-C6 *BCP* (cross) for the of the CHD molecular graph are presented, see also **Figure 1(c)** and **Figure 4**.

**Conclusions**

We have investigated the CHD→HT photoreaction using the SSR method and analyzed the subsequent density matrices with our next generation QTAIM and stress tensor methodology to determine factors influencing the CHD→HT photoreaction branching ratio. Our calculations yielded a (60:40) CHD→HT photoreaction branching ratio that is closer to the experimental value of (70:30) than results obtained from the CASSCF method. The inclusion of the dynamic electron correlation in the SSR method yielded considerable differences in the geometry of the $S_1/S_0$ MECI of CHD compared with the previously and widely used CASSCF method that does not include dynamic electron correlation. For instance, earlier CASSCF calculations yielded a five-membered ring geometry corresponding to the MECI, with the C1-C5 bond (in this work we refer to as the C4---C6 *BCP*) shorter than the C5-C6 bond. In this investigation we find that the bond-path of the C1-C5 bond i.e. the C4---C6 *BCP*, is only present in the $S_0$ state and is longer than that of the fissile C5-C6 *BCP*/C5--C6 *BCP* contrary to previous CASSCF based calculations.

Further to this, we find using the QTAIM and stress tensor analysis that the C4---C6 *BCP* provides no measurable contribution to the geometry of the $S_1/S_0$ MECI of CHD because it is very weak and topologically unstable interaction and only present with very long bond-paths > 3.8 Å. The presence of an attraction between C5 and C6 atoms even at geometries corresponding to higher relative energies ΔE implies that the C5-C6 bond is central to controlling the dynamics of CHD ring-opening. Sufficient attraction exists between the C5 and C6 atoms in the $S_0$ electronic state on the CHD side of MECI point to cause the contraction of the closed-shell C5--C6 *BCP* bond-path to temporarily revert to the shared-shell C5-C6 *BCP*. From the QTAIM *BCP* analysis the closed-shell C5--C6 *BCP* is determined to be "sticky" on the basis of values of $H(\mathbf{r_b}) < 0$ that indicate a degree of covalent character and therefore coupling to the neighboring covalent (σ-bonds) C-H *BCP*s. This indicates that the reaction can be characterized as being pulled back towards CHD. Hence, this explains the origin of the (70:30) CHD ratio as being due to a "sticky" C5--C6 *BCP*.

We also investigate the CHD→HT photoreaction using the QTAIM bond-path analysis: the bond-path framework set $\mathbb{B} = \{(\boldsymbol{p_0},\boldsymbol{p_1}),(\boldsymbol{q_0},\boldsymbol{q_1}),\boldsymbol{r}\}$ and the bond-path ellipticity ε profiles. From this we demonstrate that the strong shared-shell C5-C6 *BCP* acquires closed-shell *BCP* character early on in the reaction process and this process starts closer CHD for the $S_1$ state than for the $S_0$ state.

## Acknowledgements

The National Natural Science Foundation of China is gratefully acknowledged, project approval number: 21673071. The One Hundred Talents Foundation of Hunan Province and the aid program for the Science and Technology Innovative Research Team in Higher Educational Institutions of Hunan Province are also gratefully acknowledged for the support of S.J. and S.R.K.

# SUPPLEMENTARY MATERIALS

# Next-Generation Quantum Theory of Atoms in Molecules for the Ground and Excited State of the Ring-Opening of Cyclohexadiene (CHD)

Xin Bin[1], Tianlv Xu[1], Steven R. Kirk[1*], Michael Filatov[1,2] and Samantha Jenkins[1*]

*Key Laboratory of Chemical Biology and Traditional Chinese Medicine Research and Key Laboratory of Resource Fine-Processing and Advanced Materials of Hunan Province of MOE, College of Chemistry and Chemical Engineering, Hunan Normal University, Changsha, Hunan 410081, China*

email: steven.kirk@cantab.net
email: mike.filatov@gmail.com
email: samanthajsuman@gmail.com

**1. Supplementary Materials S1.** The variation of the eigenvector-following path length $\mathbb{H}^*$ of the shared-shell C5-C6 *BCP* with the NEB path length.

**2. Supplementary Materials S2.** The variation of $\mathbb{H}$ of the closed-shell C5-C6 *BCP* with the NEB path length.

**3. Supplementary Materials S3.** The variation $\mathbb{H}$ of the C5-C6 *BCP* with the NEB path length.

**4. Supplementary Materials S4.** Outline of the procedure to calculate $\mathbb{B}$, and discussion of the construction.

**5. Supplementary Materials S5.** The $p_0$- and $q_0$-paths for the $S_0$ state in the vicinity of the C.I.

# 1. Supplementary Materials S1.

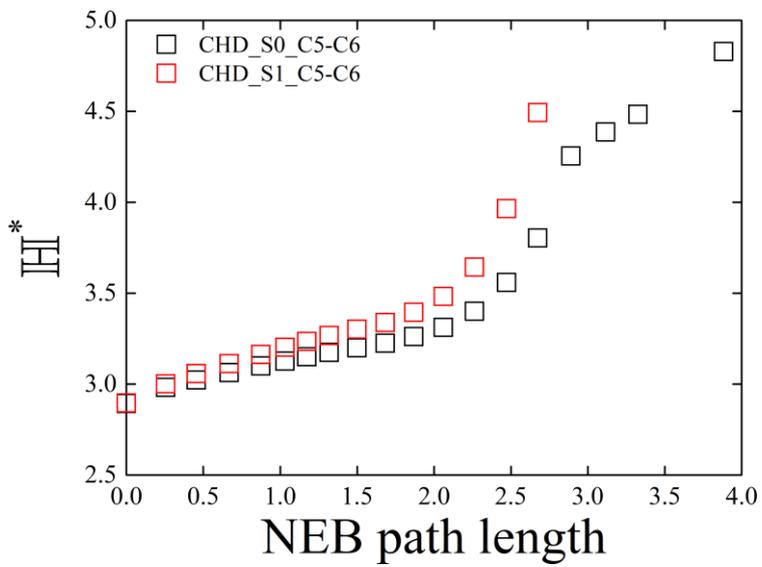

(a)

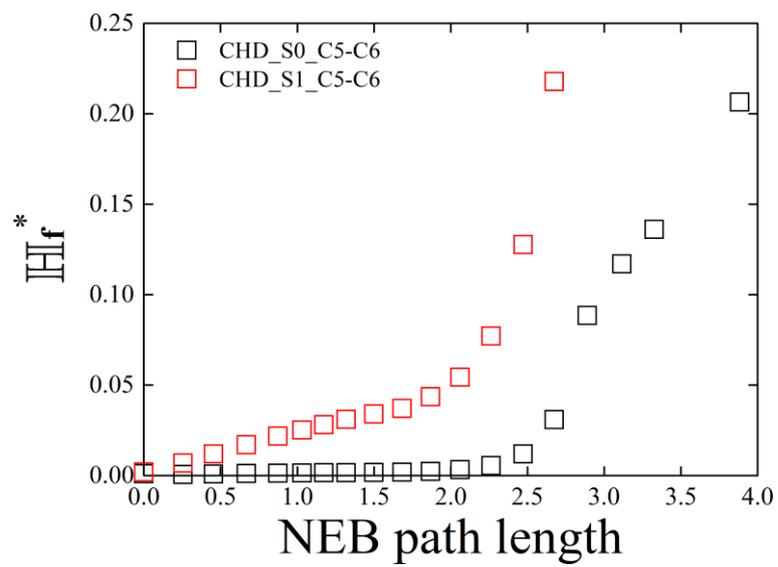

(b)

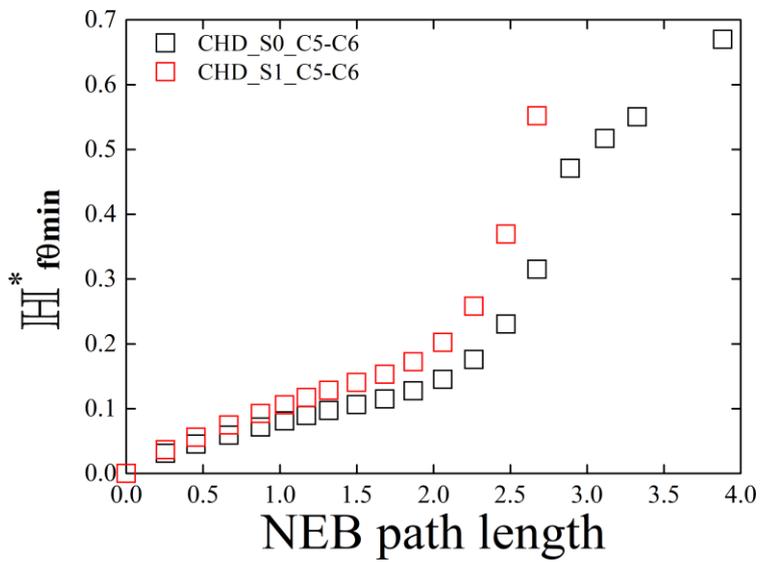

(c)

**Figure S1.** The variations of eigenvector-following path length $\mathbb{H}^*$ of the shared-shell (C5-C6 *BCP*) *BCP* are shown in (**a**). The variation of the fractional eigenvector-following path length $\mathbb{H}^*_f$ of the C5-C6 *BCP* are shown in (**b**). The variation of the $H^*_{f\theta min} = (H^* - H^*_{\theta min})/H^*_{\theta min}$ of the C5-C6 *BCP* are shown in (**c**).

## 2. Supplementary Materials S2.

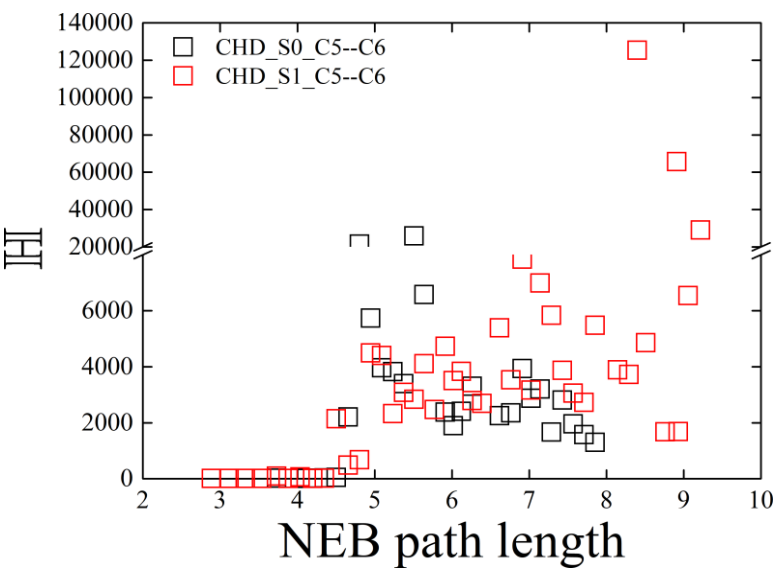

(a)

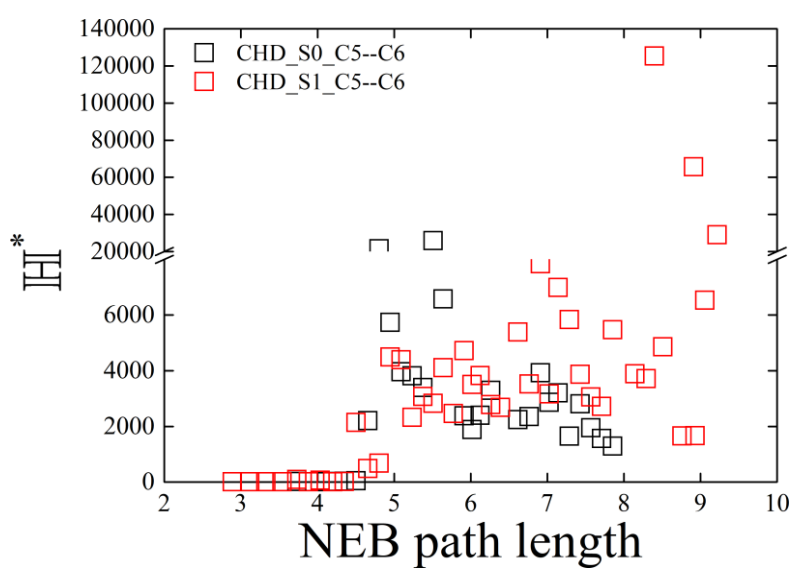

(b)

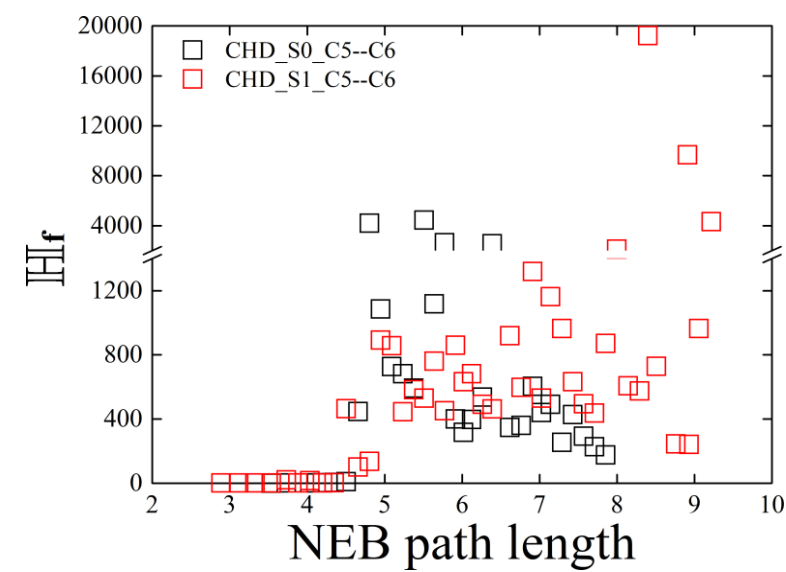

(c)

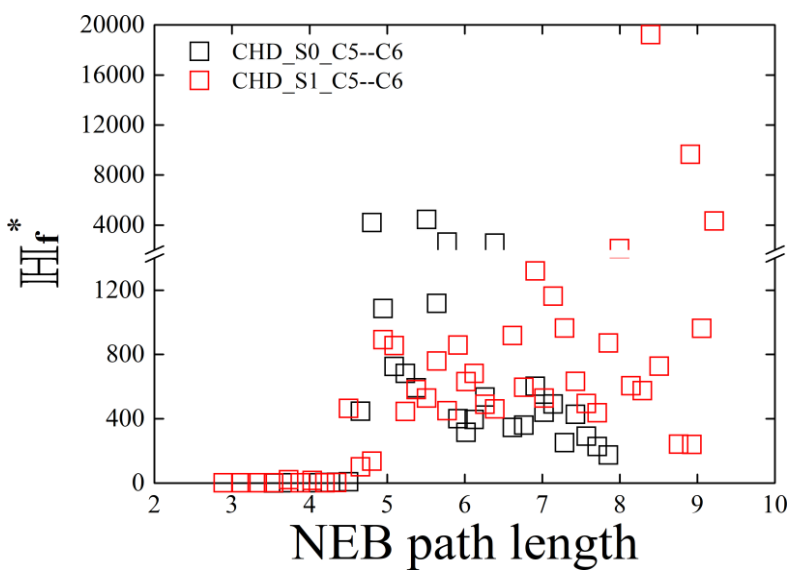

(d)

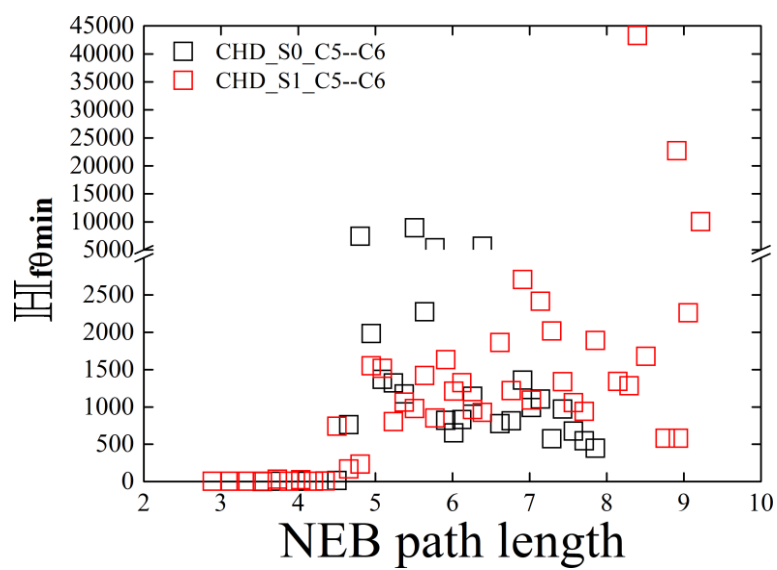

(e)

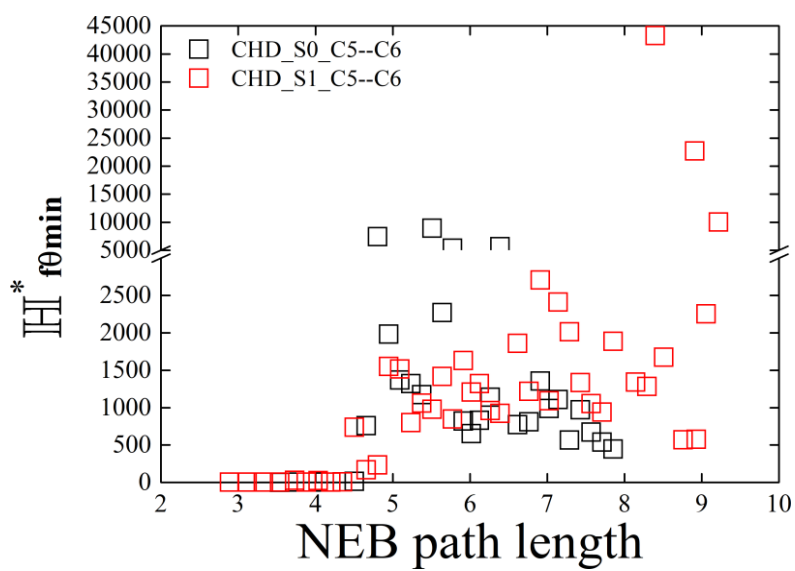

(f)

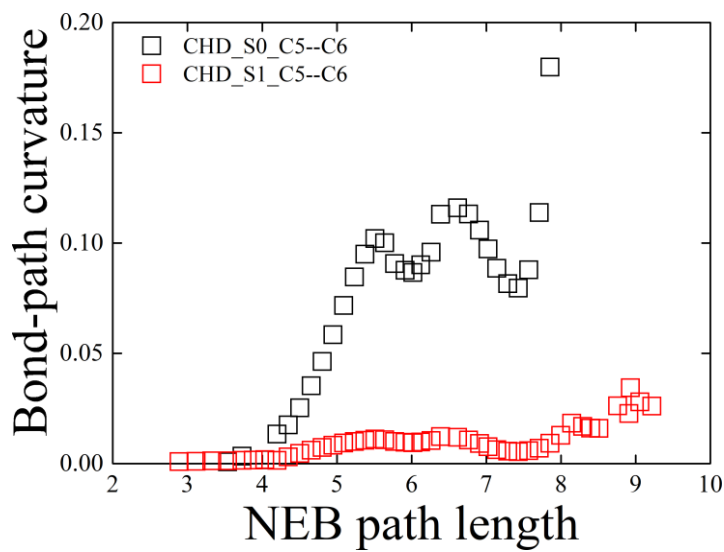
(g)

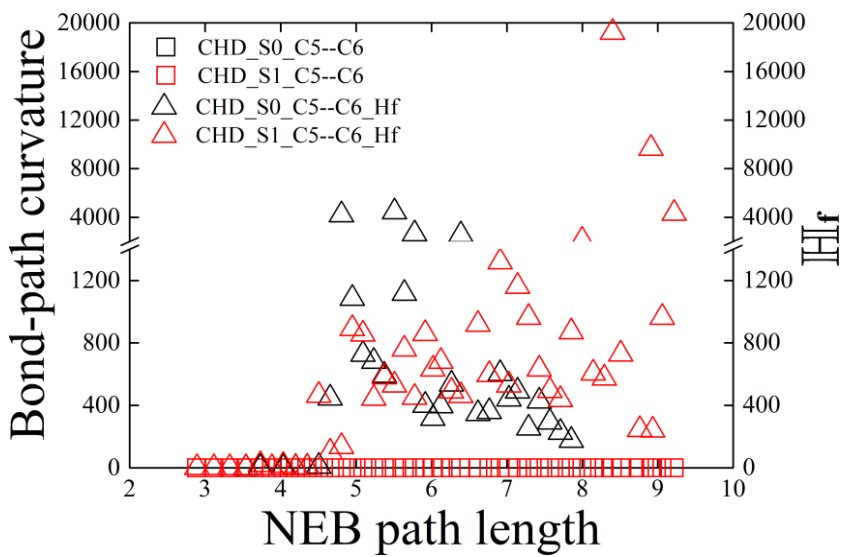
(h)

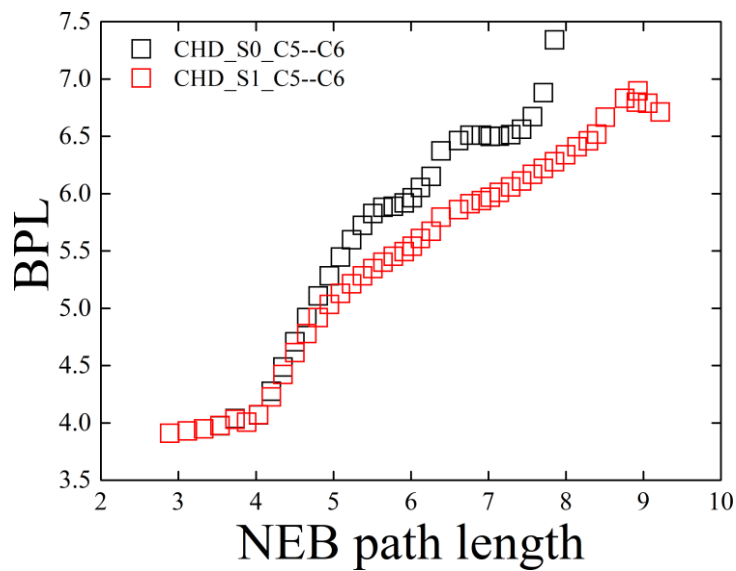
(i)

**Figure S2.** The eigenvector-following path length H and the corresponding variations eigenvector-following path length $\mathbb{H}^*$ of the closed-shell (C5-C6 *BCP*) *BCP* are shown in **(a-b)** respectively. The eigenvector-following path length $\mathbb{H}_f$ and the corresponding variation the fractional eigenvector-following path length $\mathbb{H}^*_f$ of the C5--C6 *BCP* are shown in **(c-d)** respectively. The variation of the $\mathbb{H}_{f\theta min} = (\mathbb{H} - \mathbb{H}_{\theta min})/\mathbb{H}_{\theta min}$ and the corresponding value for $\mathbb{H}^*_{f\theta min} = (\mathbb{H}^* - \mathbb{H}^*_{\theta min})/\mathbb{H}^*_{\theta min}$ of the C5--C6 *BCP* are shown in **(e-f)** respectively. The variation of the bond-path curvature (BPL-GBL)/GBL and corresponding with $\mathbb{H}_f$ of the C5--C6 *BCP* are shown in **(g-h)**. The variation of the bond-path lengths (BPL) of the C5--C6 *BCP* is shown in **(i)**.

**3. Supplementary Materials S3.**

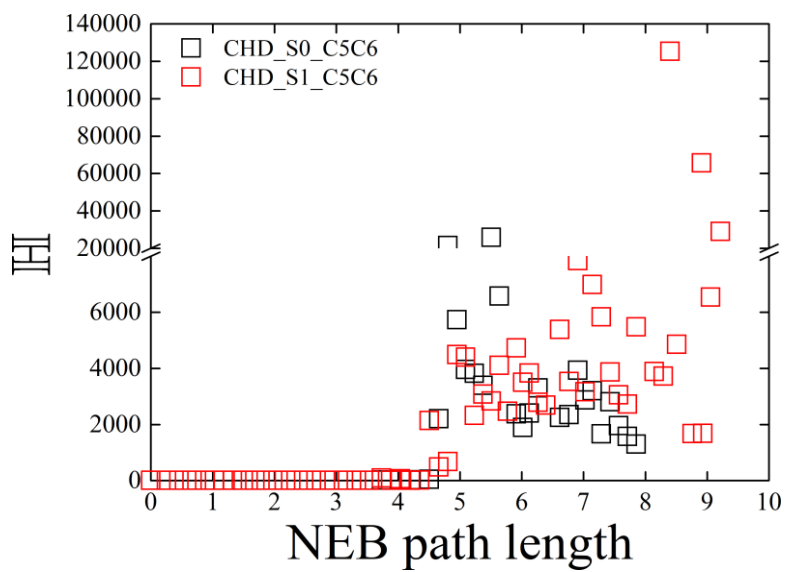

(a)

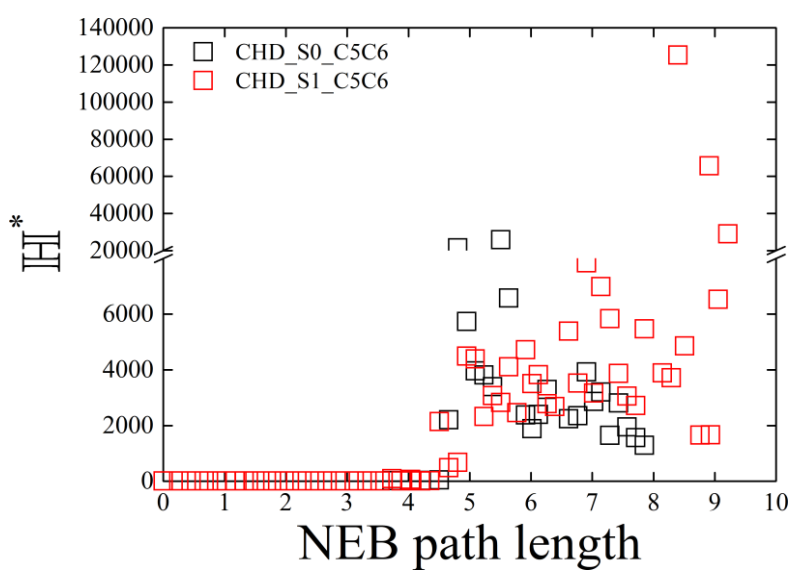

(b)

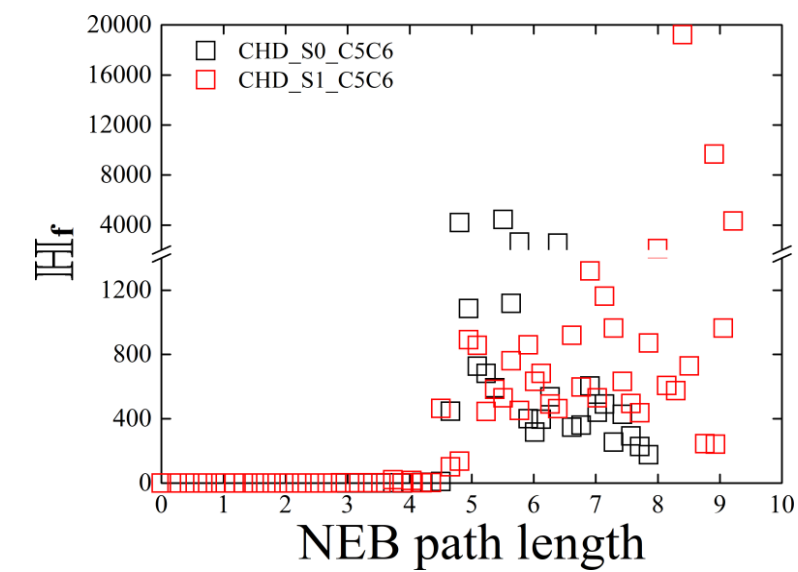

(c)

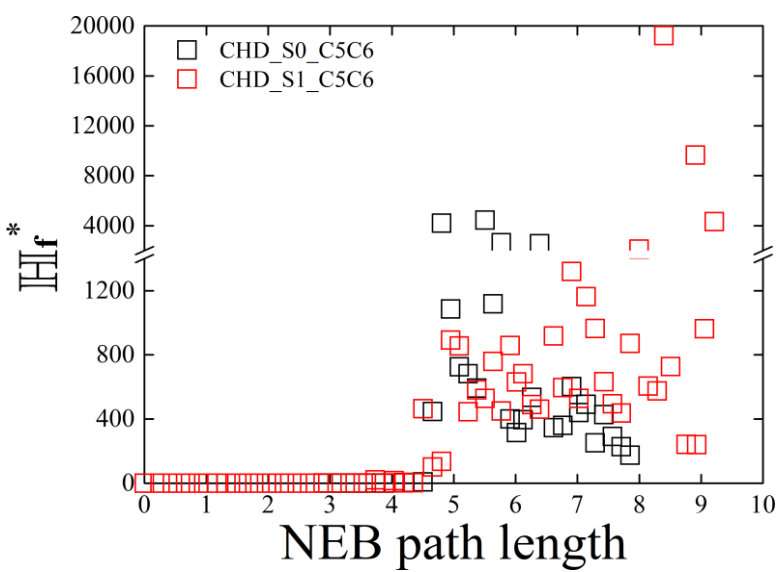

(d)

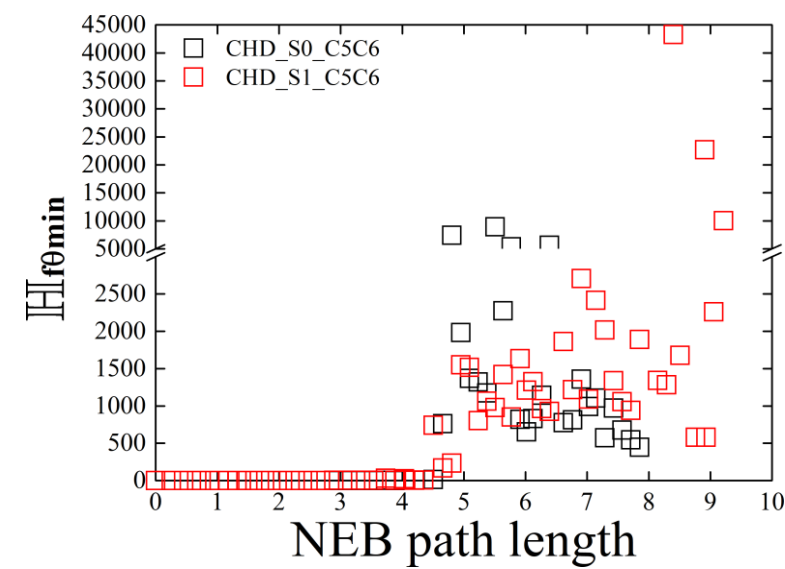

(e)

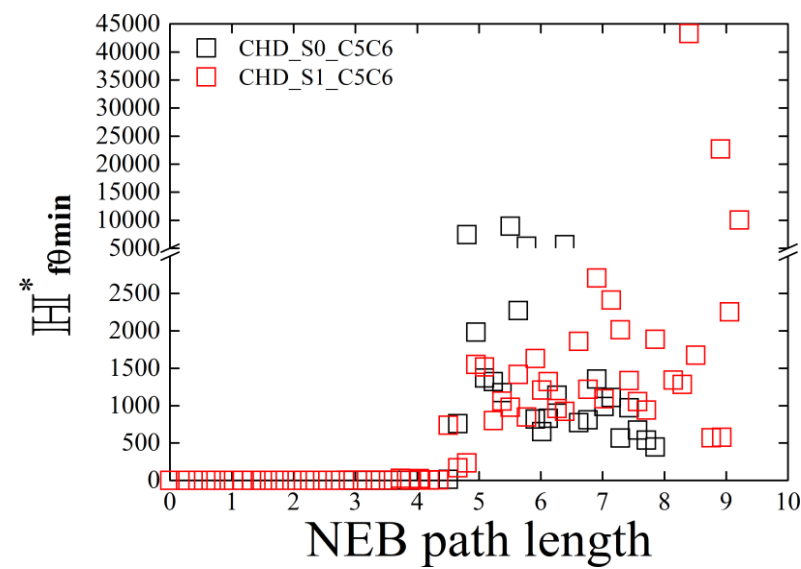

(f)

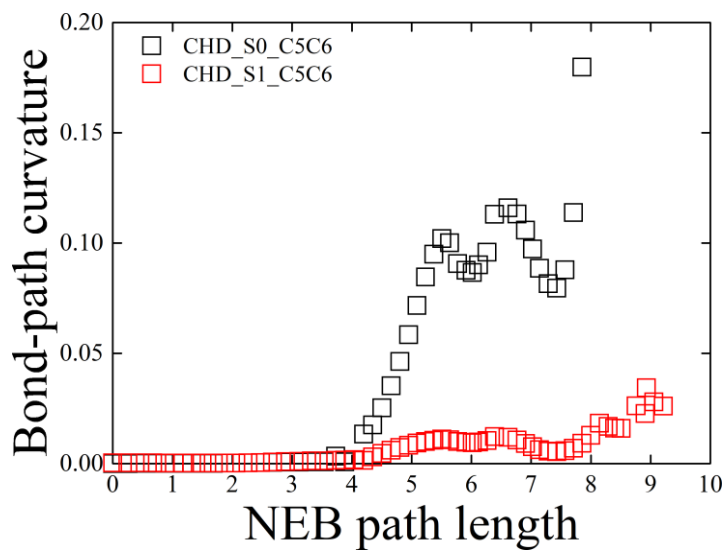
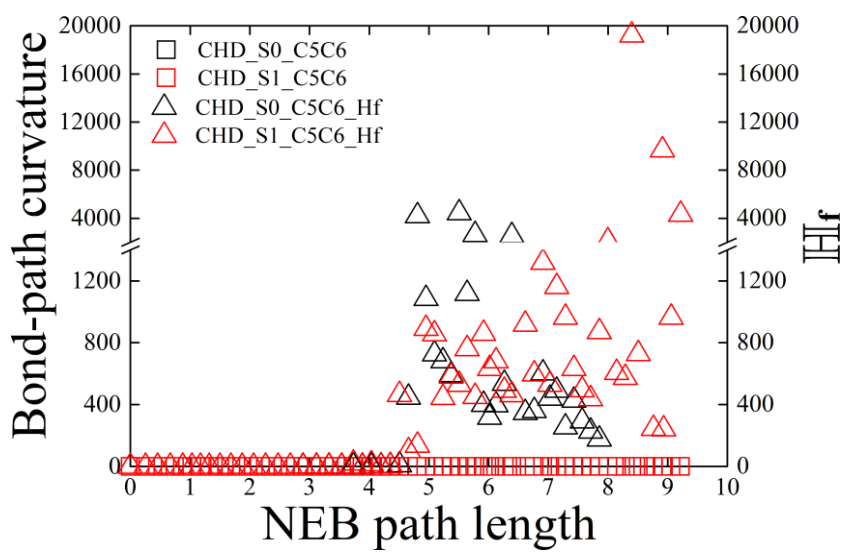

(g)

(h)

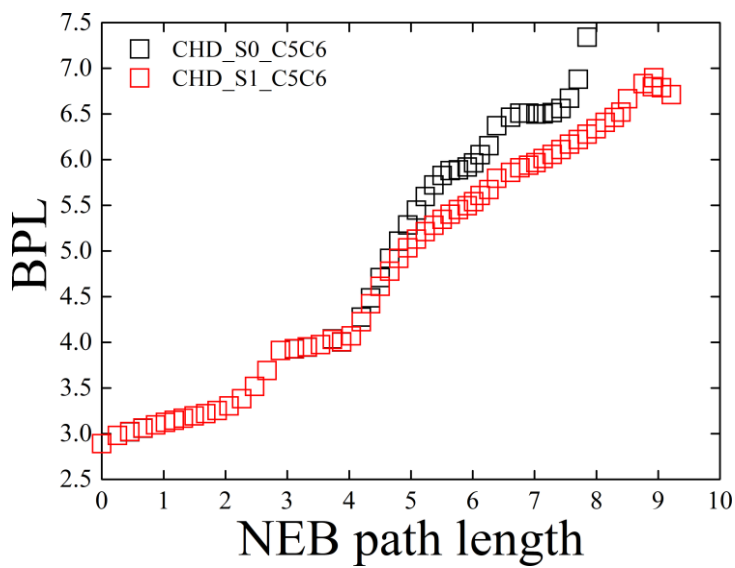

(i)

**Figure S3.** The eigenvector-following path length $\mathbb{H}$ and the corresponding variations eigenvector-following path length $\mathbb{H}^*$ of the C5-C6 *BCP* are shown in **(a-b)** respectively. The eigenvector-following path length $\mathbb{H}_f$ and the corresponding variation the fractional eigenvector-following path length $\mathbb{H}^*_f$ of the C5-C6 *BCP* are shown in **(c-d)** respectively. The variation of the $\mathbb{H}_{f\theta min} = (\mathbb{H} - \mathbb{H}_{\theta min})/\mathbb{H}_{\theta min}$ and the corresponding value for $\mathbb{H}^*_{f\theta min} = (\mathbb{H}^* - \mathbb{H}^*_{\theta min})/\mathbb{H}^*_{\theta min}$ of the C5-C6 *BCP* are shown in **(e-f)** respectively. The variation of the bond-path curvature (BPL-GBL)/GBL and corresponding with $\mathbb{H}_f$ of the C5-C6 *BCP* are shown in **(g-h)**. The variation of the bond-path lengths (BPL) of the C5-C6 *BCP* is shown in **(i)**.

## 4. Supplementary Materials S4.

**The reasons for the choice of the ellipticity ε as scaling factor.** This was motivated by the fact that the scaled vector tip paths drop smoothly onto the bond-path, ensuring that the tip paths are always continuous. We previously discussed the unsuitability of alternative scaling factors, $|\lambda_1 - \lambda_2|$ this was not pursued as it lacks the universal chemical interpretation of the ellipticity ε e.g. double-bond ε > 0.25 vs. single bond character ε ≈ 0.10. Also unsuitable choices for scaling factors, on the basis of not attaining zero, included either ratios involving the $\lambda_1$ and $\lambda_2$ eigenvalue or any inclusion of the $\lambda_3$ eigenvalue. The $\lambda_3$ eigenvalue was also found to unsuitable because it contains no information about the least ($\underline{e}_1$) and most ($\underline{e}_2$) preferred directions of the total charge density $\rho(\mathbf{r})$ accumulation.

**Discussion on the uniqueness of the $\mathbb{H}^*$ and $\mathbb{H}$.** Because $\mathbb{H}^*$ and $\mathbb{H}$ are defined by the distances swept out by the $\underline{e}_2$ tip path points $\boldsymbol{p}_i = \boldsymbol{r}_i + \varepsilon_i \underline{e}_{1,i}$ and $\boldsymbol{q}_i = \boldsymbol{r}_i + \varepsilon_i \underline{e}_{2,i}$ respectively and the scaling factor, $\varepsilon_i$ is identical in equation **(3a)** and equation **(3b)** therefore for a linear bond-path $\boldsymbol{r}$ then $\mathbb{H}^* = \mathbb{H}$. The bond-path framework set $\mathbb{B} = \{\boldsymbol{p},\boldsymbol{q},\boldsymbol{r}\}$ should consider the bond-path to comprise the *unique $\boldsymbol{p}$-, $\boldsymbol{q}$-* and *$\boldsymbol{r}$*-paths, swept out by the $\underline{e}_1$, $\underline{e}_2$ and $\underline{e}_3$, eigenvectors that form the eigenvector-following paths with lengths $\mathbb{H}^*$, $\mathbb{H}$ and BPL respectively. The $\boldsymbol{p}$- and $\boldsymbol{q}$-paths are unique even when the lengths of $\mathbb{H}^*$ and $\mathbb{H}$ are the same or very similar because the $\boldsymbol{p}$- and $\boldsymbol{q}$-paths traverse different regions of space. Bond-paths $\boldsymbol{r}$ with non-zero bond-path curvature which will result in $\mathbb{H}^*$ and $\mathbb{H}$ with different values, this is more likely to occur for the equilibrium geometries of closed-shell *BCP*s than for shared-shell *BCP*s. This is because the $\boldsymbol{p}$- and $\boldsymbol{q}$-paths will be different because of the greater distance travelled around the outside of a twisted bond-path $\boldsymbol{r}$ compared with the inside of the same twisted bond-path $\boldsymbol{r}$. This is because within QTAIM the $\underline{e}_1$, $\underline{e}_2$ and $\underline{e}_3$, eigenvectors can only be defined to within to a factor of -1, i.e. ($\underline{e}_1$,-$\underline{e}_1$), ($\underline{e}_2$,-$\underline{e}_2$) and ($\underline{e}_3$,-$\underline{e}_3$) therefore there will be two possible tip-paths. The consequences of this (within QTAIM) calculation of the $\mathbb{H}^*$ is that we dynamically update the sign convention to define $\mathbb{H}^*$ as being the shorter of the two possible tip-paths because $\underline{e}_1$ is the least preferred direction of accumulation of $\rho(\mathbf{r})$. A similar procedure is used for $\mathbb{H}$ except that we chose the longer of the two possible tip-paths because $\underline{e}_2$ is the most preferred direction of accumulation of $\rho(\mathbf{r})$.

**Implementation details of the calculation of the eigenvector-following path lengths $\mathbb{H}$ and $\mathbb{H}^*$.**

When the QTAIM eigenvectors of the Hessian of the charge density $\rho(\mathbf{r})$ are evaluated at points along the bond-path, this is done by requesting them via a spawned process which runs the selected underlying QTAIM code, which then passes the results back to the analysis code. For some datasets, it occurs that, as this evaluation considers one point after another in sequence along the bond-path, the returned calculated $\underline{e}_2$ (correspondingly $\underline{e}_1$ is used to obtain $\mathbb{H}^*$) eigenvectors can experience a 180-degree 'flip' at the 'current' bond-path point compared with those evaluated at both the 'previous' and 'next' bond-path points in the sequence. These 'flipped' $\underline{e}_2$ (or $\underline{e}_1$) eigenvectors, caused by the underlying details of the numerical implementation in the code that computed them, are perfectly valid, as these are defined to within a scale factor of -1 (i.e. inversion). The analysis code used in this work detects and re-inverts such temporary 'flips' in the $\underline{e}_2$ (or $\underline{e}_1$) eigenvectors to maintain consistency with the calculated $\underline{e}_2$ (or $\underline{e}_1$) eigenvectors at neighboring bond-path points, in the evaluation of eigenvector-following path lengths $\mathbb{H}$ and $\mathbb{H}^*$.

## 5. Supplementary Materials S5.

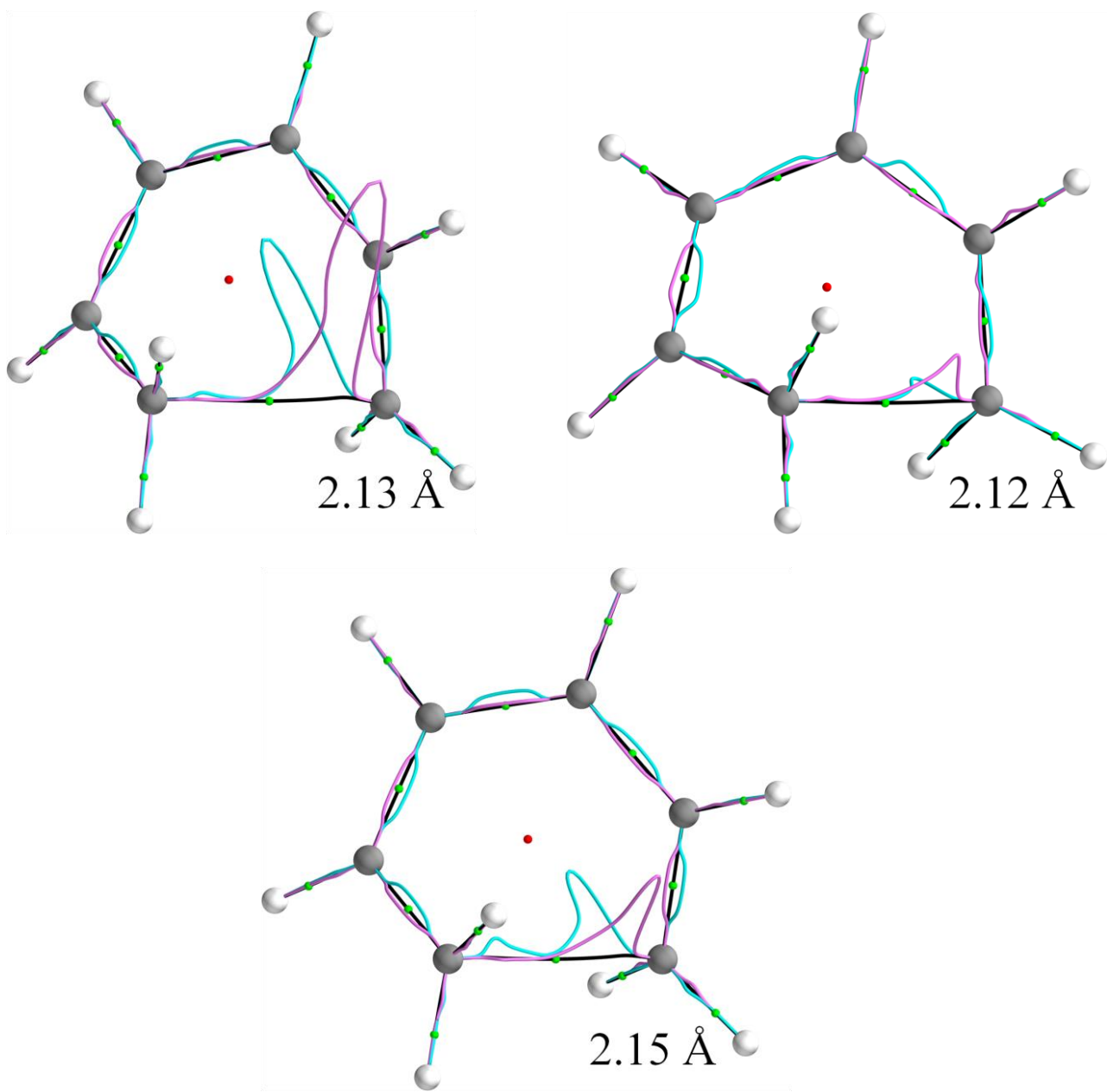

**Figure S5.** The values of $\mathbb{B} = \{(p_0,q_0),r\}$ corresponding to the $S_0$ state of the CHD→HT photoreaction for values of R(C5-C6) = 2.13 Å, 2.12 Å and 2.15 Å are presented in sub-figures **(a)-(c)** respectively. Note that these values of R(C5-C6) correspond to the closed-shell C5--C6 *BCP*, shared-shell C5-C6 *BCP* and closed-shell C5--C6 *BCP* bond-path in sub-figures **(a)-(c)** respectively.